\documentclass[prb,showpacs,groupedaddress,twocolumn]{revtex4}
\usepackage{amsmath}
\usepackage{mathrsfs}
\usepackage{txfonts}
\usepackage{amssymb}
\usepackage{graphicx}
\usepackage{hyperref}
\usepackage{ulem}
 \usepackage{overpic}
 \usepackage{psfrag}

\begin{document}

\title{Non-local correlations in the Haldane phase for an XXZ
      spin-1 chain: A perspective from infinite matrix product state representation}

\author{Yao Heng Su}
\affiliation{Center for Modern Physics and Department of Physics,
 Chongqing University, Chongqing 400044, The People's Republic of
 China}

\author{Sam Young Cho}
 \email{sycho@cqu.edu.cn}
 \affiliation{Center for Modern Physics and Department of Physics,
 Chongqing University, Chongqing 400044, The People's Republic of
 China}

\author{Bo Li}
\affiliation{Center for Modern Physics and Department of Physics,
 Chongqing University, Chongqing 400044, The People's Republic of
 China}

\author{Hong-Lei Wang}
\affiliation{Center for Modern Physics and Department of Physics,
 Chongqing University, Chongqing 400044, The People's Republic of
 China}

\author{Huan-Qiang Zhou}
\affiliation{Center for Modern Physics and Department of Physics,
 Chongqing University, Chongqing 400044, The People's Republic of
 China}

\begin{abstract}
  String correlations are investigated in an infinite-size XXZ spin-1 chain.
  By using the infinite matrix product state representation,
  we calculate a long-range string order directly
  rather than an extrapolated string
  order in a finite-size system.
  In the N\'eel phase, the string correlations decay
  exponentially. In the XY phase (Tomonaga-Luttinger liquid phase),
  the behaviors of the string correlations show
  a unique two-step decaying to zero within a relatively very large
  lattice distance,
  which makes a finite-size study difficult to verify
  the non-existence of the string order.
  Thus, in the Haldane phase,
  the non-vanishing string correlations in the limit of a very large distance
  allow to characterize the phase boundaries to the XY phase and
  the N\'eel phase, which implies that
  the transverse long-range string order is the order parameter for the Haldane phase.
  In addition, the singular behaviors of the von Neumann entropy
  and the fidelity per lattice site
  are shown to capture clearly the phase transition points that are
  consistent with the results from the transverse long-range string
  order.
  The estimated critical points including a Berezinsky-Kosterlitz-Thouless
  transition
  from the XY phase to the Haldane phase
  agree well with
  the previous results:
  $\Delta_{c2} = 0$ for the XY-Haldane phase transition
  and $\Delta_{c3} = 1.185$ for the Haldane-N\'eel phase transition
  from the density renormalization group.
  From a finite-entanglement scaling of the von Neumann entropy
  with respect to  the truncation dimension,
  the central charges are found to be
  $c \simeq 1.0$ at  $\Delta_{c2} = 0$ and $c \simeq 0.5$ at $\Delta_{c3} = 1.185$,
  respectively, which shows that
  the XY-Haldane  phase transition at $\Delta_{c2}=0$
  belongs to the Heisenberg universality class,
  while the Haldane-N\'eel phase transition at $\Delta_{c2}=1.185$
  belongs to the two-dimensional classical Ising universality class.
  It is also shown that, the long-range order parameters
  and the von Neumann entropy, as well as
  the fidelity per site approach, can
  be applied to characterize quantum phase transitions as a
  universal phase transition
  indicator for one-dimensional lattice many-body systems.
\end{abstract}

\date{\today}

\pacs{75.10.Pq, 75.40.Cx, 75.40.Mg, 03.67.Mn}

\maketitle

\section{Introduction}

 Since Landau introducd
 the theory of second order phase transitions,
 understanding local order
 parameters characterizing different quantum phases
 has become
 one of the main paradigms
 in condensed matter physics \cite {Landau,QPT}.
 Local order parameters are then known to detect a spontaneous symmetry breaking
 for quantum phase transitions \cite{Belitz,Chaikin}.
 In some cases, however,
 this local order parameter approach does not work because
 quantum phase transitions arise from the cooperative behavior of a
 system
 due to emergence of non-local order \cite{Hatsugai}.
 Although a non-local order parameter could not be directly probed in experiments,
 it can be useful for understanding the underlying physics of quantum phases
 as well as for marking phase boundaries.
 Thus, understanding non-local orders in low-dimensional spin
 systems at absolute zero
 temperature have been an important subject in quantum phase
 transitions
 in recent years
 \cite{Kitazawa1,Schollw,Kennedy1,Kennedy2,Garlea,Charrier,Alcaraz1}.

 A prototype example, in particular,
 is the spin-1 antiferromagnetic Heisenberg chain \cite{Schulz,Alcaraz2}.
 The groundstate of the spin chain is in a distinct phase
 with a finite energy gap, but does not exhibit any local order parameter \cite{Kennedy1}.
 These properties of the
 groundstate are comprehensively understood by
 a non-vanishing non-local string correlation \cite{Tasaki} introduced
 by den Nijs and Rommelse \cite{Nijs}, and by understanding the Affleck-Kennedy-Lieb-Tasaki
 (AKLT) Hamiltonians for interger-spin chains \cite{Affleck,Kolezhuk}.
 The distinct phase of the groundstate is called the Haldane phase \cite{Haldane}.
 The energy gap is also called the Haldane gap, which has been manifested by
 experimental evidences found in
 CsNiCl$_3$ \cite{Buyers} and the organic crystal
 Ni(C$_2$H$_8$N$_2$)$_2$NO$_2$ClO$_4$ \cite{Katsumata}.

 Indeed,
 in order to characterize the Haldane phase,
 such non-local string correlations have been extensively studied in
 various finite-size spin systems such as anisotropic spin-1 Heisenberg chains \cite{Alcaraz2},
 frustrated antiferromagnetic Heisenberg spin-1 chains \cite{Kolezhuk},
 alternating Heisenberg chains \cite{Hida,Yamamoto},
 spin ladders \cite{White1,Shelton,Anfuso} and tubes \cite{Garlea},
 restricted solid-on-solid model \cite{Nijs}, lattice boson
 systems \cite{Berg}, and so on.
%
%
%
%
 The density matrix renormalization group (DMRG) \cite{Sakai},
 the large-cluster-decomposition Monte Carlo method \cite{Nomura}, and
 exact diagonalization with Lanczos method \cite{Yajima,Botet} have been applied
 for these studies.
 In a recently developed tensor network (TN) representation, i.e,
 matrix product state (MPS) representation \cite{mps},
 the DMRG method \cite{Ueda} also
 has been applied to explore a string correlation for a finite-size lattice.
 The non-local string order
 inferred from the string correlation behaviors
 in such finite-size spin systems has been used to characterize the Haldane
 phase from other phases \cite{Ueda,Alcaraz2}.
%
%
 In fact, it is then believed that string correlations can characterize the Haldane
 phase. However, no characterization of the Haldane phase, to the best of our knowledge,
 has been made by directly computing
 long-range string order (LRSO) instead of
 the extrapolated behaviors of string correlations till now because all pervious studies
 have been carried out in finite-size systems.

 Thus, in this study, we will investigate string correlations and
 their
 extreme values for very large spin lattices, i.e., directly computing the string order.
 To do this, we consider the infinite-size spin-1 antiferromagnetic Heisenberg chain
 with anisotropic exchange interaction $\Delta$.
 We will employ
 the infinite matrix product state (iMPS) representation \cite{Vidal1,Vidal2,Zhao1} for the ground
 state wavefunction of the infinite lattice system.
 The groundstate wavefunction can be obtained numerically by using
 the infinite time evolving block decimation (iTEBD) method \cite{Vidal2}
 within the iMPS representation.
 To investigate string correlations,
 we will introduce an efficient way to calculate a non-local correlation and
 its extreme value for a large-size lattice system in the iMPS representation.
 It is found that, except for the Haldane phase, the string correlations decay
 exponentially in the N\'eel phase, while they show
 a unique behavior of decaying to zero within very large lattice distance
 in the XY phase (Tomonaga-Luttinger liquid phase).
 For the Haldane phase,
 the string correlations are saturated
 to finite values, which shows a LRSO as the lattice distance
 goes to infinity.
 Also,
 from the LRSO with respect to the anisotropic interaction
 parameter $\Delta$, it is clearly shown that \textit{both the $x$- and
 $y$-components  rather than the $z$-component of the LRSO
 play a role as the order parameters characterizing the Haldane phase.}
 As a consequence,
 the string order parameters enable us to directly characterize
 the possible phases of the system with respect to the anisotropic exchange
 interaction.
 Moreover, the von Neumann entropy and the fidelity per lattice site (FLS)
 are calculated to show that their singular
 behaviors correspond to the phase transition points.
 The central charges from the finite-entanglement scaling
 quantify the universality classes of the transition points.
 The FLS is shown to capture a Berezinsky-Kosterlitz-Thouless (BKT) type transition,
 in contrast to
 the fidelity susceptibility that fails to detect it.


 This paper is organized as follows.
 In Sec. II, a brief explanation for the iMPS
 representation is given.
 We discuss how to capture non-local correlations
 including string correlations and string orders
 directly by exploiting the iTEBD method.
 In Sec. III, the spin-1 XXZ chain model is
 introduced.
 We discuss the behaviors of the
 string correlations and N\'eel correlations as a function
 of the lattice distance for given anisotropic
 interaction strengths in Sev. IV.
 The phase diagram of the spin-1 XXZ chain model is presented based on
 the non-local correlations and the string and N\'eel order parameters in Sec. V.
 In Sec. VI, we discuss local and non-local properties
 of the iMPS groundstate that allow to introduce pseudo symmetry breaking order for the XY
 phase for finite truncation dimensions.
 In Sec. VII,
 the phase transitions and their universality classes
 are discussed from the von Neumann entropy and the central charges via
 the finite-entanglement scaling.
 In Sec. VIII, the groundstate
 FLS is shown to have a clear pinch point
 that corresponds to a quantum phase transition.
 In Sec. IX, our conclusions and remarks are given.

\section{iMPS representation and non-local correlations in numerical method}
 Recently, significant progress has been made in
 numerical studies based on TN representations
 \cite{mps,Ueda,Vidal1,Vidal2,Murg,Zhao1,Wang1,DMRG,Verstraete,Li}
 for the investigation of quantum phase transitions,
 which offers a new perspective from quantum entanglement and fidelity, thus
 providing a deeper understanding on characterizing critical
 phenomena
 in finite and infinite spin lattice systems.
 Actually, a wave function represented in
 TNs allows to perform the classical
 simulation of quantum many-body systems.
 Especially, in one-dimensional spin systems,
 a wave function for infinite-size lattices
 can be described by the iMPS representation \cite{mps}.
 The iMPS representation
 have been successfully applied to investigate
 the properties of ground-state wave functions in various infinite spin lattice
 systems.
 The examples include Ising model in a transverse magnetic field \cite{Zhao1}
 and with antisymmetric anisotropic and alternative bond interactions\cite{Li},
 XYX model in an external magnetic field \cite{Zhao1},
 and spin-1/2 XXZ model \cite{Wang1}.
 However, the iMPS has not been applied much to explore spin
 correlations.
 Few studies have shown the
 behaviors of spin-spin correlations in the infinite Ising spin chain \cite{Vidal2}.
 Furthermore,
 non-local spin correlations have not been explored yet in infinite-size
 systems with the iMPS representation.
 Then, in this section, we will discuss how to calculate
 a non-local spin correlation within the iMPS representation.

\subsection{iMPS representation and iTEBD algorithm}

\begin{figure}
\includegraphics[angle=-90,width=3.2in]{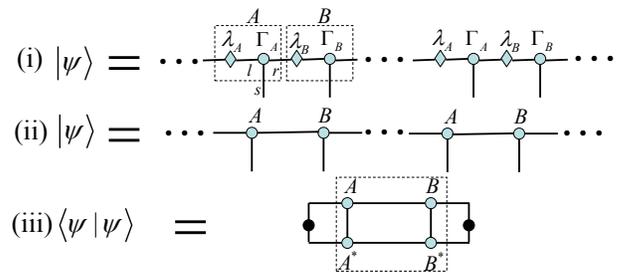}
\caption{(color online)
 (i) Diagrammatic infinite matrix product state (iMPS) representation of
  a wavefunction $|\Psi\rangle$
 having two-site translational
 invariance
 for infinite one-dimensional lattice
 systems. $\lambda_{A(B)}$ denoted by filled diamonds
 are a diagonal (singular value) matrix,
 respectively, depending on the $2i$-th ($A$) and $2i+1$-th ($B$) bonds.
 $\Gamma_{A(B)}$ indicated by filled circles
 are a three-index tensor for the $2i$-th and $2i+1$-th
 sites, respectively. $r$ and $l$ indicate the bond indices.
 (ii) A compact form of the pictorial representation in (i), where
 $A$ and $B$ (filled circles)  by
 absorbing $\lambda$ into the tensors $\Gamma$ (indicated by the
 dashed-line
 box in (i)) present the lattice sites.
(iii) The norm $\langle \Psi | \Psi \rangle$ of a given state
 $|\Psi\rangle$ in (i).
 The black dots indicate the left and right dominant
 eigenvectors that can be determined for the dominant eigenvalue of the transfer matrix $T$.
 The transfer matrix $T$ is obtained by the contraction on
 the tensors $A$, $A^*$, $B$, and $B^*$ in
 the dashed-line box.
 } \label{fig1}
\end{figure}

 For an infinite one-dimensional lattice system,
 a state can be written as \cite{Vidal2,Wolf}

 \begin{eqnarray}
 &&|\Psi\rangle
  =  
  \sum_{\{S\}} \sum_{\{\alpha\}}
 \cdots
 \lambda^{[i]}_{\alpha_i} \Gamma^{[i]}_{\alpha_i,s_i,\alpha_{i+1}}
 \lambda^{[i+1]}_{\alpha_{i+1}} \Gamma^{[i+1]}_{\alpha_{i+1},s_{i+1},\alpha_{i+2}}
 \lambda^{[i+2]}_{\alpha_{i+2}}
\cdots
 \nonumber
 \\
  && ~\hspace*{2.0cm} \times |\cdots S_{i-1}S_{i}S_{i+1}
 \cdots \rangle,
 \label{wave}
 \end{eqnarray}
 where $|S_i\rangle$ denote a basis of the local Hilbert space at
 the site $i$, the elements of a diagonal matrix $\lambda^{[i]}_{\alpha_i}$
 are the Schmidt decomposition coefficients of the bipartition
 between the semi-infinite chains $L(-\infty,...,i)$ and
 $R(i+1,...,\infty)$,
 and $\Gamma^{[i]}_{\alpha_i,S_i,\alpha_{i+1}}$ are a three-index
 tensor. The physical indices $S_i$ take the value $1, \cdots, d$
 with the local Hilbert space dimension $d$ at the site $i$.
 The bond indices $\alpha_i$ take the value $1, \cdots, \chi$
 with the truncation dimension of the local Hilbert space at the
 site $i$. The bond indices connect the tensors $\Gamma$ in the nearest neighbor sites.
 Such a representation in Eq. (\ref{wave}) is called
 the iMPS representation \cite{Vidal2}.
 If a system Hamiltonian has a translational invariance, one can
 introduce a translational invariant iMPS representation for a state.
 Practically, for instance, for a two-site translational invariance,
 the state can be reexpressed in terms of only the three-index tensors
 $\Gamma_{A(B)}$ and the two diagonal matrices $\lambda_{A(B)}$ for
 the even (odd) sites \cite{Li}, where $\{\Gamma, \lambda\}$ are in the canonical form,
 i.e.,
 \begin{eqnarray}
 |\Psi\rangle
  =  
  \sum_{\{S\}} \sum_{\{l,r\}}
 \cdots
 \lambda_A \Gamma_A
 \lambda_B \Gamma_B
 \lambda_A
 \cdots
  |\cdots S_{i-1}S_{i}S_{i+1}
 \cdots \rangle,
 \label{state2}
 \end{eqnarray}
 where $l$ and $r$ are the left and right bond indices, respectively.
 In Fig. \ref{fig1} (i), a state $|\Psi\rangle$  with a two-site translational
 invariance is pictorially displayed
 in the iMPS representation
 for infinite one-dimensional lattice systems.
 In a more compact form, further, the quantum state can be reexpressed
 as the state in Fig. \ref{fig1}
(ii) by absorbing the diagonal matrices $\lambda$ into the tensors
 $\Gamma$.

 Once a random initial state $|\Psi(0)\rangle$ is prepared in the iMPS representation,
 one may employ the iTEBD
 algorithm \cite{Vidal2} to calculate a groundstate wavefunction numerically.
 For instance, if a system Hamiltonian is
 translational invariant and the interaction between spins consists of
 the nearest-neighbor interactions, i.e.,
 the Hamiltonian can be expressed by $H=\sum_{i}h^{[i,i+1]}$, where $h^{[i,i+1]}$
 is the nearest-neighbor two-body Hamiltonian density,
 a groundstate  wavefunction of the system can be expressed in the
 form in Eq. (\ref{state2}).
 The imaginary
 time evolution of the prepared initial state $|\Psi(0)\rangle$, i.e.,
\begin{equation}
|\Psi(\tau)\rangle=\frac{\exp[-H\tau]|\Psi(0)\rangle}{||\exp[-H\tau]|\Psi(0)\rangle||},
\end{equation}
 leads to a groundstate of the system for a large enough $\tau$.
 By using the Suzuki-Trotter decomposition \cite{suzuki},
 actually, the imaginary time evolution
 operator $U=\exp[-H\tau]$
 can be reduced to a product of two-site evolution operators $U(i,i+1)$ that only
 acts on two successive sites $i$ and $i+1$.
 For the numerical imaginary time evolution operation,
 the continuous time evolution can be approximately realized by
 a sequence of the time slice evolution gates $U(i,i+1)=\exp\left[-h^{[i,i+1]}
 \delta\tau\right]$
 for the imaginary time slice $\delta \tau = \tau/n \ll 1$.
 A time-slice evolution gate operation
 contracts $\Gamma_A$, $\Gamma_B$, one $\lambda_A$, two $\lambda_B$,
 and the evolution operator $U(i,i+1)=\exp\left[-h^{[i,i+1]}
 \delta\tau\right]$.
 In order to recover the evolved state in the iMPS representation,
 a singular value decomposition (SVD) is performed
 and the $\chi$ largest singular values are obtained.
 From the SVD,
 the new tensors $\Gamma_A$, $\Gamma_B$, and $\lambda_A$
 are generated. The latter is used to update the tensors $\lambda_A$ as the new one for all other sites.
 Similar contraction on the new tensors
 $\Gamma_A$, $\Gamma_B$, two new $\lambda_A$, one  $\lambda_B$,
 and the evolution operator $U(i+1,i+2)=\exp\left[-h^{[i+1,i+2]}\right]$,
 and its SVD produce the updated $\Gamma_A$, $\Gamma_B$, and $\lambda_B$
 for all other sites.
 After the time-slice evolution, then,
 all the tensors $\Gamma_A$, $\Gamma_B$, $\lambda_A$, and $\lambda_B$
 are updated.
 This procedure is repeatedly performed until the system energy
 converges to a groundstate energy that yields a groundstate
 wavefunction in the iMPS representation.
 The normalization of the groundstate wavefunction is guaranteed by
 requiring the norm $\langle \Psi | \Psi \rangle = 1$ in Fig. \ref{fig1} (iii).

\begin{figure}
\includegraphics[angle=-90,width=3.4in]{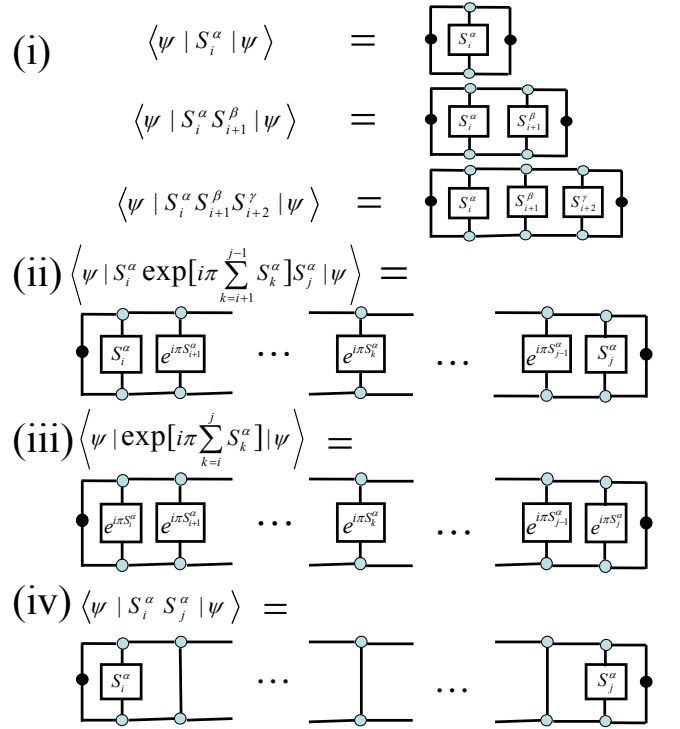}
\caption{(color online)
 Diagrammatic representations of the expectation values in the iMPS representation.
 The
 black dots in left (right) side denote the dominant eigenvectors of
 the transfer matrix. In between a given wave
 function $|\Psi\rangle$ and its complex conjugate, each spin
 operator act on each site.
 (i) Local order computations including one, two, and three spins such as
  magnetization, dimer order, and chiral orders.
 Non-local correlations can also be calculated in the iMPS representation,
 e.g., (ii) string correlation
 $O^\alpha_\textrm{S} (i,j)=- \langle S^\alpha_i \exp [i\pi
 \sum_{k=i+1}^{j-1}S^\alpha _k ]S^\alpha _j \rangle\, (\alpha \in \{x, y,
 z\})$,
 (iii) parity correlation
 $O^\alpha_\textrm{P} (i,j)= \langle \exp [i\pi
 \sum_{k=i}^{j}S^\alpha _k ] \rangle$,
 and
 (iv) N\'{e}el
 correlation (two point spin correlation) $O^\alpha_\textrm{N} (i,j)=
 (-1)^{i-j} \langle S^\alpha_i S^\alpha _j \rangle.$
 }
 \label{fig2}
\end{figure}

\subsection{Non-local correlations}
 In principle, once one obtains a groundstate wavefunction,
 the expectation values
 of physical quantities can be calculated.
 In Fig. \ref{fig2},
 we depict the diagrammatic iMPS representations for some examples
 of various expectation value calculations.
 Figure \ref{fig2} (i) presents
 the computation of successive spin operators such as
 magnetization \cite{Zhao1} $\langle S^\alpha_i \rangle$, dimer order \cite{Yu} $\langle S^\alpha_i S^\beta_{i+1}
 \rangle$,
 and chiral order \cite{Selinger,Hikihara} $\langle S^\alpha_i S^\beta_{i+1} S^\gamma_{i+2}
 \rangle$ ($\alpha, \beta, \gamma \in \{ x, y, z\} $).
 On calculating the expectation values,
 each spin operator acting on a site
 is sandwiched between a given wave function $|\Psi\rangle$ and its complex conjugate.
 The left and right dominant eigenvectors of
 the transfer matrix denoted by the black dots
 act on the tensor, for instance, contracting the tensors $A$,
 $A^*$, (or $B$, $B^*$) and the local spin operator $S_i$ for $\langle S^\alpha_i \rangle$.
 This leads to the expectation value $\langle S^\alpha_i \rangle$.

 This procedure can be simply expanded to the calculation of
 spin-spin correlations as well as non-local correlations.
 Examples are the string correlation ${\cal O}_S$ \cite{Nijs},
 the parity correlation ${\cal O}_P$ \cite{Hikihara}, and
 the N\'eel correlation ${\cal O}_N$ \cite{Yu} given by, respectively,
\begin{subequations}
\begin{eqnarray}
 {\cal O}^{\, \alpha}_\textrm{S} (i,j)
 &=& -\Bigg \langle S^\alpha_i \exp \Bigg[i\pi
      \sum_{k=i+1}^{j-1}S^\alpha _k \Bigg]S^\alpha _j \Bigg \rangle,
 \\
 {\cal O}^{\, \alpha}_\textrm{P} (i,j)
 &=& \Bigg \langle \exp \Bigg[i\pi
       \sum_{k=i}^{j} S^\alpha_k \Bigg] \Bigg \rangle,
 \\
 {\cal O}^{\, \alpha}_\textrm{N} (i,j)
 &=& (-1)^{i-j}\langle S^\alpha_i S^\alpha _j
\rangle,
\end{eqnarray}
\end{subequations}
 where $i$ and $j$ denote the site locations in the lattice
 and then the lattice distance is $|i-j|$.
 Figures \ref{fig2} (ii), (iii), and (iv) present
 the computation of
 the string, parity, and N\'eel correlations, respectively, in the iMPS
 representation.
 Note that, for the calculation of these correlations,
 the $(i-j)$ tensors are involved.
 For the parity correlation, the multi-site operator
 $\exp[i\pi \sum S^\alpha _k]$ acts on each site between
 the site $i$ and site $j$.
 For the string correlations, the spin operators $S_i$ and $S_j$
 act on sites $i$ and $j$ while the multi-site operator
 $\exp[i\pi \sum S^\alpha _k]$ acts on the sites in between
 the sites $i$ and $j$.
 Compared to the string correlations,
 the N\'eel correlations can be calculated by replacing
 the multi-site operator
 $\exp[i\pi \sum S^\alpha _k]$ with the identity operator, i.e.,
 $I_{i+1}\cdots I_{j-1}$ while
 the parity correlation can be obtained by expanding the multi-site spin operator to
 the both ends of the lattice distance.
 Then,
 it should be noted that, in principle, \textit{ the iMPS representation allows to
 calculate any correlations in the limit of the infinite distance, i.e.,
 $|i-j| \rightarrow \infty$}.
 For numerical calculations, in order to obtain the correlations in the limit
 of the infinite distance,
 one can set a truncation error $\varepsilon$ rather than the lattice distance,
 i.e., ${\cal O}^{\, \alpha} (i,j)-{\cal O}^{\, \alpha}
 (i,j+1) < \varepsilon$. In this study, for instance, $\varepsilon =
 10^{-8}$ is chosen.

\begin{figure}
\includegraphics[width=3.3in]{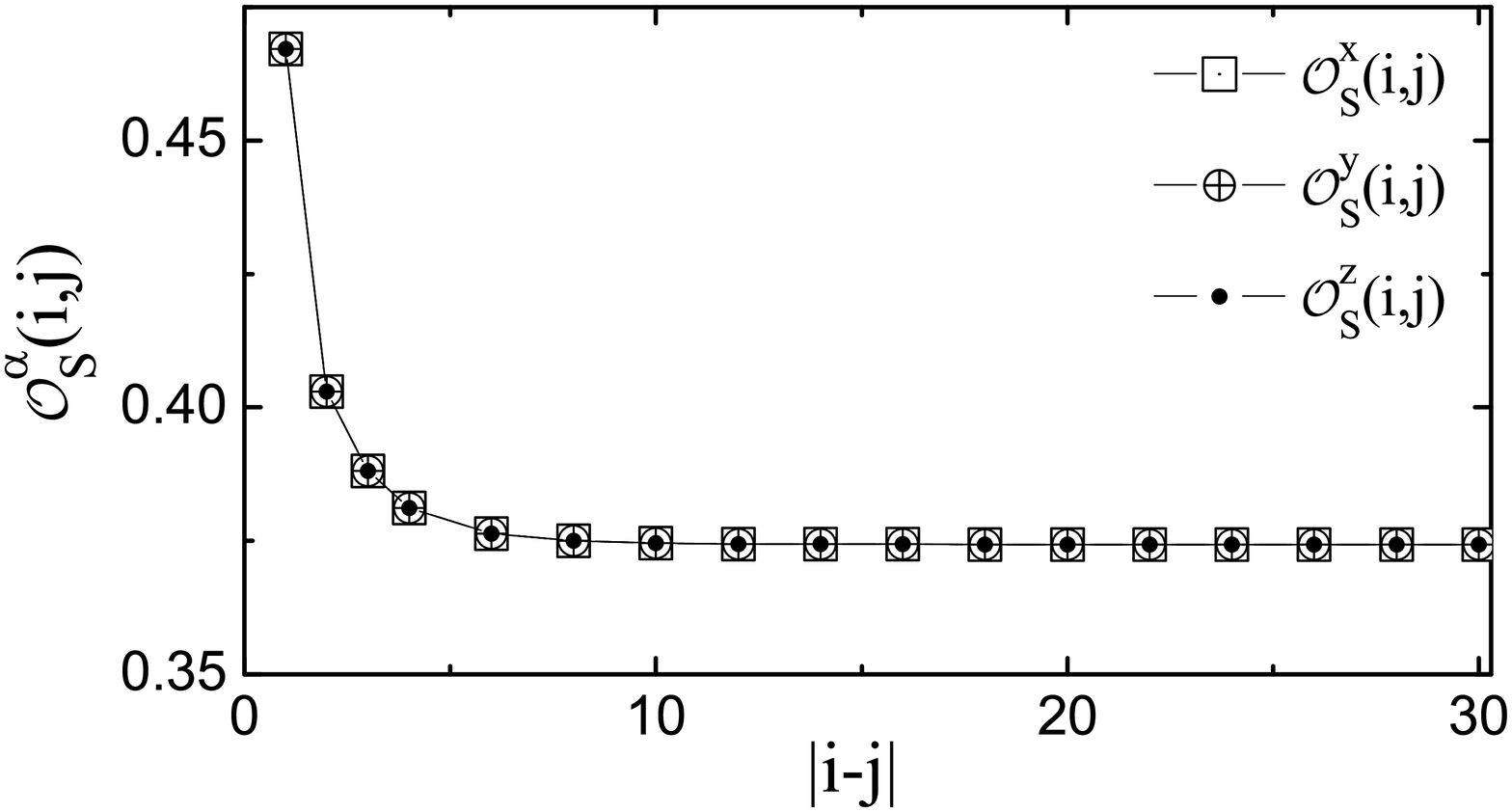}
\caption{(color online)
 String correlation ${\cal O}^\alpha_{S} (i,j)$ for the Haldane spin-1
 chain with truncation dimension $\chi=32$ in the iMPS representation.
 All components
 of the string correlations are well saturated to the value
 $O^\alpha_{S} = 0.37434447$, which agrees very well with the result
 from the DMRG in Ref. \onlinecite{White}.
 }
\label{fig3}
\end{figure}

 As an example,
 for the spin-1 Haldane chain
 $H = \sum_{i} \mathbf{S}_i \cdot
 \mathbf{S}_{i+1}$,
 the exact-diagonalization calculations of a 14-site lattice have
 estimated
 ${\cal O}^{\, z}_{\textrm{S}}(\infty) \simeq 0.38$ \cite{Girvin}.
 The DMRG methods
 have verified the existence of LRSO by estimating
 ${\cal O}^{\, z}_{\textrm{S}}(\infty) = 0.37432509$
 for the Haldane chain \cite{White}.
 In Fig. \ref{fig3}, we plot the string correlation
 ${\cal O}^{\,\alpha}_{\textrm{S}} (i,j)$
 as a function of the lattice distance $|i-j|$ with the truncation dimension
 $\chi=32$.
 It is shown clearly that the ${\cal O}^{\,\alpha}_{\textrm{S}} (i,j)$ starts to saturate
 around the lattice distance $|i-j| \simeq 10$ and
 ${\cal O}^{\,x}_{\textrm{S}} (i,j)={\cal O}^{\,y}_{\textrm{S}} (i,j)={\cal O}^{\,z}_{\textrm{S}} (i,j)$
 for the Haldane chain.
 The saturated value of the string correlation from our iMPS
 representation approach
 is given as ${\cal O}^\alpha_{\textrm{S}}(\infty) = 0.37434447$,
 which agrees very well with the value
 ${\cal O}^z_{\textrm{S}}(\infty) = 0.37432509$
 as well as the saturation behavior of the string correlations
 from the DMRG method in Fig. 5 of Ref.
 \onlinecite{White}.
 As is well-known, further,
 the spin-1 AKLT model
 $H^{S=1}_{AKLT} = \sum_{i} [(1/2) \mathbf{S}_i \cdot
 \mathbf{S}_{i+1} + (1/6)(\mathbf{S}_i \cdot
 \mathbf{S}_{i+1})^2 + 1/3]$
 is exactly solvable and
 the string order is given as the exact value
 $4/9$ \cite{Affleck,Kolezhuk}. In our iMPS representation, the
 value of the string order has been confirmed to be $4/9$
 for the spin-1 AKLT model within the machine accuracy.
%


\section{spin-1 $XXZ$ Heisenberg chain}
 Spin-1 Heisenberg chains are one of the prototypical examples in understanding non-local
 correlations \cite{Alcaraz1}, i.e., string correlation.
 Then, to investigate non-local correlations in one-dimensional spin systems,
 we consider an infinite spin-1 $XXZ$ Heisenberg chain
 described by the Hamiltonian
\begin{equation}\label{Hamiltonian}
 H = J\sum^{\infty}_{i=-\infty}
     [S^x_iS^x_{i+1}+S^y_iS^y_{i+1}+\Delta S^z_iS^z_{i+1}],
\end{equation}
 where $S^\alpha_i(\alpha=x, y, z)$ are the spin-1 operators at the
 lattice site $i$, $J$ denotes the antiferromagnetic spin-exchange
 interaction
 between the nearest neighbor spins, and
 $\Delta$ is responsible for the anisotropy of the exchange interaction.
 This model has been intensively studied for a couple of decades
 \cite{Hatsugai,Kennedy1,Kennedy2,
 Ueda,wei,Murashima,Pan,Alcaraz1,
 Botet,Nomura,Sakai,Yajima,Totsuka,
 Tonegawa,Yamanaka,Boschi,Ren,Gu1}.
 The studies have shown that there are the four characteristic phases
 with respect to
 the anisotropic exchange interaction $\Delta$.
 If the anisotropic interaction is much smaller than $-1$, i.e., $\Delta \ll -1$,
 the Hamiltonian can be reduced to a spin-1 ferromagnetic Ising model
 $H \approx - \sum_{i}
      S^z_i S^z_{i+1}$ and the system is in the ferromagnetic phase.
 At
 $\Delta_{c1}=-1$, a first-order transition occurs between the ferromagnetic phase
 and the XY phase.
 If the anisotropic interaction is much greater than $1$, i.e., $\Delta \gg 1$,
 the Hamiltonian is reduced to a spin-1 antiferromagnetic Ising model
 $H \approx \sum_{i} S^z_i S^z_{i+1}$ and the system is in the antiferromagnetic (AF) phase.
 At $\Delta_{c3}=1.17\pm0.02$,
 the Haldane-N\'eel phase transition occurs, which belongs
 to the two-dimensional Ising universality class \cite{Nomura,Sakai}.
 In between the two phase transition points $\Delta_{c1} < \Delta < \Delta_{c3}$,
 the XY-Haldane phase transition occurs, which has been thought to be a
 BKT transition at
 $\Delta_{c2}$ \cite{Alcaraz1,Botet,Nomura,Sakai}.

 As is well-known, many antiferromagnetic spin systems can be explored
 by the standard spin-spin correlations (N\'eel
 correlations) \cite{Yu}. However, in the Haldane phase,
 the spin-spin correlations decay exponentially with a finite
 correlation length \cite{Lajko,White,Yamanaka,Ueda} and the Haldane gap exists.
 In this aspect, the Haldane phase can be considered as a disordered phase
 \cite{Charrier}.
 Also, the XY phase is characterized by the power
 law decay of the spin-spin correlations (N\'eel correlations) \cite{Lajko} with gapless
 excitations.
 Characterizing both the XY and Haldane phases therefore is a non-trivial task
 in the aspect of the spin-spin correlations.
 By investigating the spin correlations,
 the transition point $\Delta_{c2}$ has been estimated
 to be $0\lesssim \Delta_{c2}\lesssim 0.2$ from the exact numerical calculations
 and
 the finite-cell-scaling analysis \cite{Botet},
 $\Delta_{c2}=-0.01\pm0.03$ from the phenomenological renormalization-group
 technique and the finite-size scaling analysis with 16 spin sites \cite{Sakai},
 and $\Delta_{c2}=0.068 \pm 0.003$ from the criterion
 exponents of the spin correlations $\eta_x=1/4$ in the exact
 diagonalization method with 16 spin sites \cite{Yajima}.
 Investigating the excitation gap, as the anisotropic interaction strength
 varies, is also a method to characterize the Haldane phase.
 By using the
 lowest-energy levels and finite-size scaling for energy gaps from the Lanczos
 method, the critical point has been
 conjectured to be at $\Delta_{c2} = 0$ in Ref. \onlinecite{Alcaraz1}.
 As an alternative way to characterize the Haldane phase,
 the string order arising due to the fully broken
 $Z_2 \times Z_2$ hidden symmetry
 has been investigated \cite{Kennedy1,Kennedy2}.
 The XY-Haldane transition point has been estimated
 $\Delta_{c2} \sim 0$ by exploring the string correlations from a finite size analysis
 \cite{Alcaraz2}
 and by a finite-size scaling of the string order \cite{Ueda}.

\section{Behaviors of the string and N\'eel correlations
    in  spin-1 XXZ chain}

\begin{figure}
\includegraphics[width=3.3in]{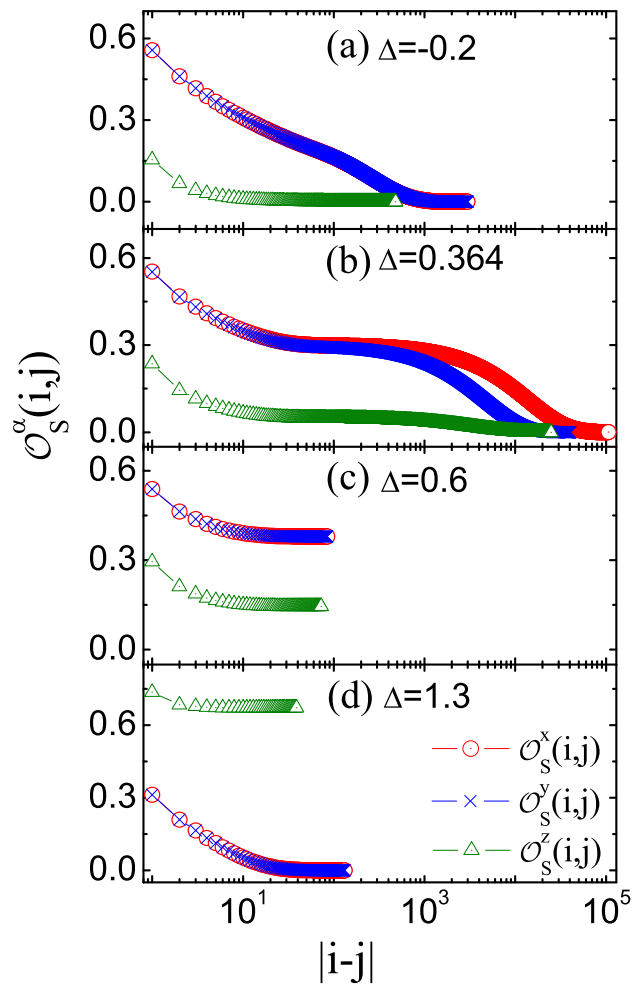}
\caption{(color online)
 String correlations ${\cal O}^\alpha_{S}$ as
 a function of the lattice distance $|i-j|$ with the truncation
 dimension $\chi =32$
for various
 anisotropic interaction
 (a) $\Delta=-0.2$ and
 (b) $\Delta= 0.364$ in the XY phase, (c) $\Delta=0.6$ in the Haldane phase,
 and (d) $\Delta=1.3$ in the N\'eel phase.
 In fact, for the truncation dimension
 $\chi=32$, the phase transition points are given as
 $\Delta_{c2}(\chi=32)=0.366$ and
 $\Delta_{c3}(\chi=32)=1.180$ in Fig. \ref{fig6}.
 It should be noted that, in the XY phase, the string correlations
 decay to zero with a unique two-step decaying behavior.
}
 \label{fig4}
\end{figure}

 A non-vanishing correlation in the limit of the infinite
 lattice distance ($|i-j|\rightarrow \infty$), i.e.,
 a long-rang order reveals that the system is in a ordered state.
 For the non-local correlations,
 the string and N\'eel orders are respectively defined by
\begin{subequations}
\begin{eqnarray}
 O^{\, \alpha}_\textrm{S} &=& \lim_{|i-j|\rightarrow \infty}
            {\cal O}^{\, \alpha}_\textrm{S} (i,j),
            \\
 O^{\, \alpha} _\textrm{N} &=& \lim_{|i-j|\rightarrow \infty}
            {\cal O}^{\, \alpha}_\textrm{N} (i,j).
\end{eqnarray}
\end{subequations}
 For instance, the non-vanishing spin-spin (N\'eel) correlations
 for $|i-j|\rightarrow \infty$
 indicate that the system is in an antiferromagnetic state.
 Also, the ground state in the
 Haldane phase is known to be characterized by the string order \cite{Kennedy2}.
 In the viewpoint of the string order, then,
 the Haldane phase could be an ordered phase. Further, if the string
 order plays a role as the order parameter for the
 Haldane phase, from the string order,
 the phase transition boundary from the Haldane phase to
 other phases can be captured.
 This view has been applied to the investigations
 of the Haldane phase of spin-1 systems \cite{Alcaraz2,Boschi,Totsuka,Tonegawa}. By using
 numerical-diagonalization, however, available system sizes were
 too small to convince string order behaviors
 as an order parameter.
 Thus, in Ref. \onlinecite{Ueda}, comparisons between
 the behaviors of the string and the N\'eel correlations
 from a finite size spin lattice (up to 300 sites),
 and their finite-size scaling behaviors have been
 used to capture the phase boundary.
 However, directly capturing the critical behavior of the string order near the
 transition point was quite difficult due to a very limited
 lattice size.
 Compared to such approaches, as discussed in Sec. II,
 the iMPS approach enables to explore the behaviors of the string
 order directly in the limit of the infinite
 lattice distance ($|i-j|\rightarrow \infty$).

 In Fig. \ref{fig4},
 we plot the string correlations
 ${\cal O}^{\, \alpha}_{S} (i,j)$ as a function of $|i-j|$ for various
 anisotropic interactions $\Delta$.
 In Figs. \ref{fig4} (c) for the Haldane phase
 and (d) for the N\'eel (antiferromagnetic) phase,
 the string correlations show a logarithmical decaying to its saturated value
 or zero
 as the lattice distance $|i-j|$ increases up to a few hundreds.
 While, in the XY phase in Figs. \ref{fig4} (a) and (b) ,
 the string correlations show a \textit{unique
 two-step decaying to zero}.
 As the lattice distance $|i-j|$ increases, that is,
 the string correlations undergo a decaying behavior
 for a few hundreds of the lattice distance, a saturation-like behavior
 for a few thousands of the lattice distance,
 and then eventually decaying again down to zero around a few tens of
 thousands.
 Hence, in contrast to the Haldane phase,
 \textit{there is no long-range string order in the XY phase.}
 Also, it should be noted that, near the transition point in the XY phase
 in Fig. \ref{fig4} (b),
 such saturation-like behaviors of the string correlations
 occur for a very wide range of the lattice distance
 from a few hundreds to a few thousands, i.e,
 roughly $10^{2}  \lesssim  |i-j|  \lesssim 10^{4}$.
 While, away from the transition point in Fig. \ref{fig4} (a),
 such saturation-like behaviors of the string correlations
 occur for a relatively narrow range of the lattice distance
 roughly $2 \times 10^{2}  \lesssim  |i-j|  \lesssim 10^{3}$.
 This saturation-like behavior of the string correlations
 makes the characterization of the Haldane phase
 quite difficult directly from a string order behavior in finite size systems.
 Indeed, compared to our iMPS results,
 a finite size system in the XY phase
 has given a finite value of the string correlation ${\cal O}^{\alpha}_{S} (i,j)$
 within the limitation of the finite system sizes (up to 300 sites) \cite{Ueda}.

\begin{figure}
\includegraphics[width=3.3in]{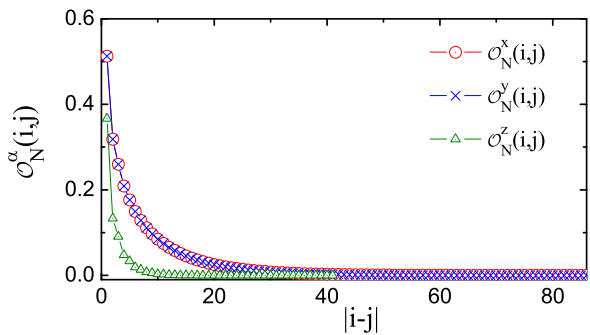}
\caption{(color online)
  N\'{e}el correlations ${\cal O}^\alpha_\textrm{N}(i,j)$
  in the Haldane phase ($\Delta=0.8$) with truncation dimension
 $\chi=32$.
 All components of the N\'eel correlations decay to zero exponentially.
 The transverse ($x$- and $y$- components)  N\'{e}el correlations
 have
 a same value, i.e., ${\cal O}^{\, x}_N={\cal O}^{\, y}_N$. }
\label{fig5}
\end{figure}
%
 In Fig. \ref{fig5}, a N\'eel correlation is displayed
 as a function of the lattice distance $|i-j|$.
 It is shown that,
 in the Haldane phase $\Delta=0.8$,
 the spin correlations (N\'eel
 correlations) decay exponentially to zero,
 which allows to characterize the phase transition
 from the N\'eel phase to the Haldane phase.
 As shown in Fig. \ref{fig4} (d),
 in the N\'eel phase, in contrast to the $z$-component of
 the string correlations that survives for very large distances,
 the $x$- and $y$-components of the string
 correlations also decay exponentially to zero.
 Also, as shown in Fig. \ref{fig4} (c),
 all the components of the string correlations in the Haldane
 phase have non-zero values in the limit of the infinite lattice
 distance.
 Then, alternatively, the $x$- and $y$-components of the string
 order
 make it possible to distinguish the Haldane phase from the N\'eel phase.
 Further, in the XY phase in Fig. \ref{fig4} (b),
 all the components of the string order become zero with the unique
 two-step decaying behavior. As a consequence,
 \textit{the $x$- and $y$-components of the string order
 play a role of a true order parameter characterizing the Haldane
 phase from the XY phase and the N\'eel phase.}

\section{Haldane phase and order parameter}

\begin{figure}
\includegraphics[width=3.3in]{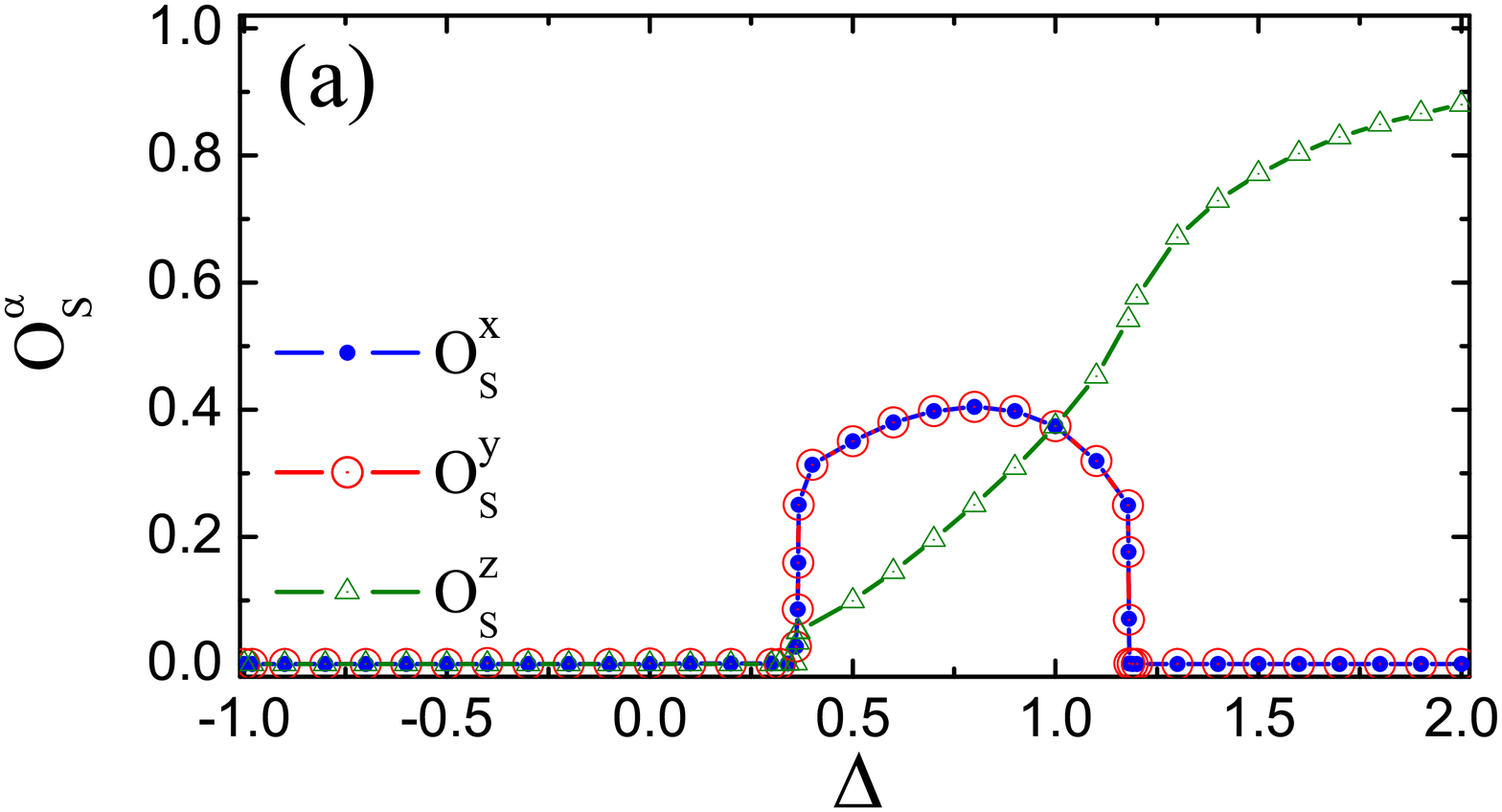}
\includegraphics[width=3.3in]{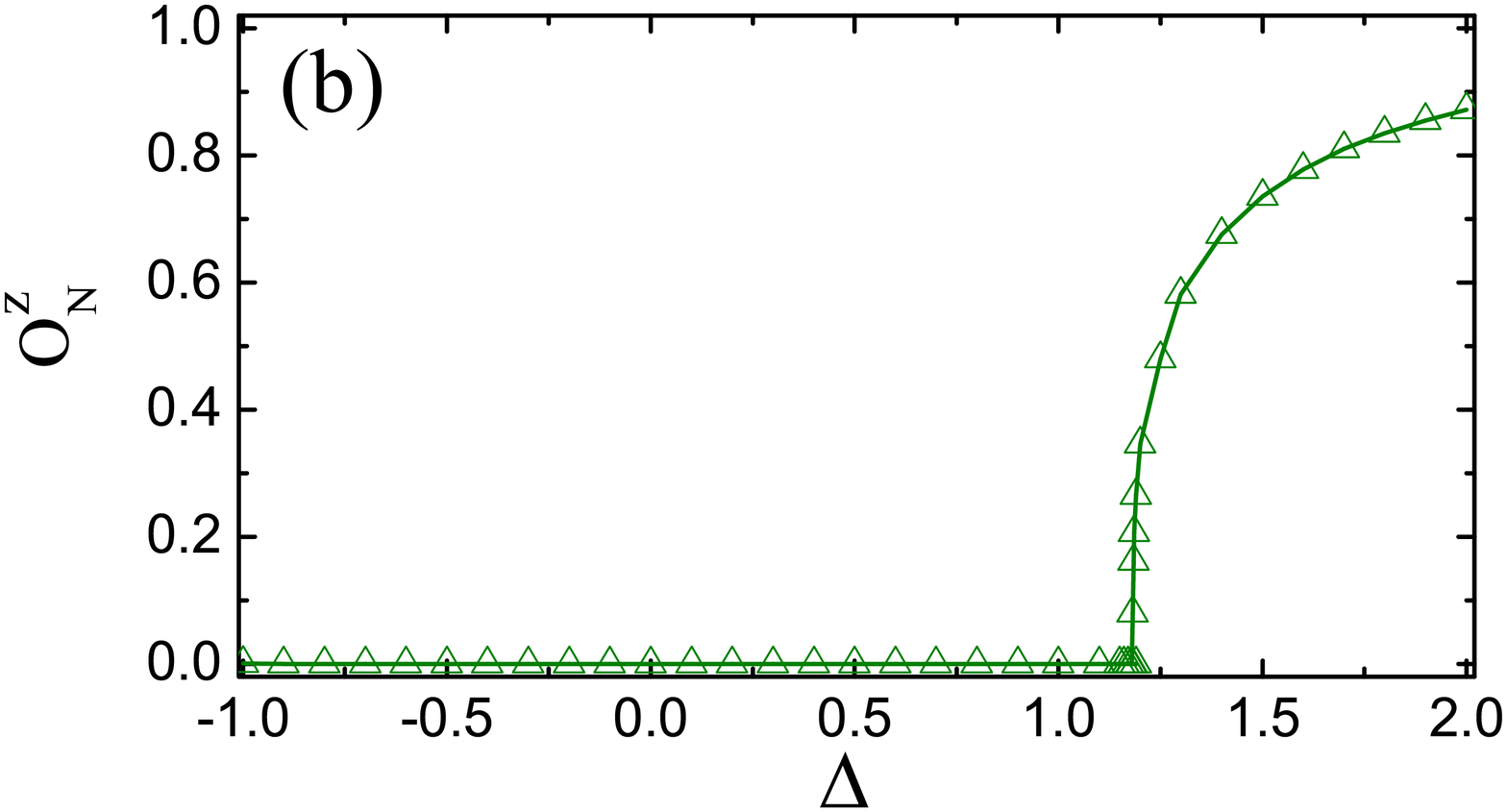}
\caption{(color online)
 (a) String order parameters ${O}_{S}^{\, x, y}$ for the
 Haldane phase
 and (b) N\'eel order parameter ${O}^{\, z}_{N}$ for the N\'eel phase as
 a function of $\Delta$ with truncation dimension $\chi =32$.
 Note that the string order ${O}_{S}^{\, z}$ does not play
 a role as the order parameter to characterize the Haldane phase in (a).
 For this truncation dimension $\chi=32$, the system is in
 the XY phase for $-1 < \Delta < 0.366$,
 the Haldane phase for $0.366 <  \Delta < 1.180$,
 and the N\'eel phase for $\Delta > 1.180$.
 For the truncation dimension $\chi =32$, then,
 the transition points are $\Delta_{c2}=0.366$
 and $\Delta_{c3} = 1.180$.
 }
 \label{fig6}
\end{figure}

 As discussed in Sec. IV, the LRSOs can characterize the
 Haldane phase. In Fig. \ref{fig6}, we plot (a) the string orders ${\cal O}^{\,
 \alpha}_{S}$
 and (b) the N\'eel order parameter ${O}^{\, z}_{N}$ as a function of the
 isotropic exchange interaction strength $\Delta$
 for the truncation dimension $\chi = 32$.
 It is shown that
 the ${O}^{\, x}_{S}={O}^{\, y}_{S}$ have non-zero values
 for $0.366 <  \Delta < 1.180$,
 while the ${O}^{\, z}_{S}$ has a finite value for $\Delta > 1.180$.
 The string orders become zero for $ -1 < \Delta < 0.366$.
 Also, the N\'eel order parameter ${O}^{\, z}_{N}$
 has a non-zero value for $\Delta > 1.180$, which characterizes the N\'eel phase.
 This implies that the ${O}^{\, x}_{S}$ and ${O}^{\, y}_{S}$
 are the order parameters for the Haldane phase.
 Then, the Haldane phase exists in the range of the anisotropic interaction
 strength $0.366 < \Delta < 1.180$, which implies $\Delta_{c2}=0.366$
 and $\Delta_{c3} = 1.180$ for the truncation dimension $\chi =32$.
 Thus,
 the XY phase occurs for $ -1 < \Delta < 0.366$.

\begin{figure}
\vspace{0.5cm}
\includegraphics[width=3.3in]{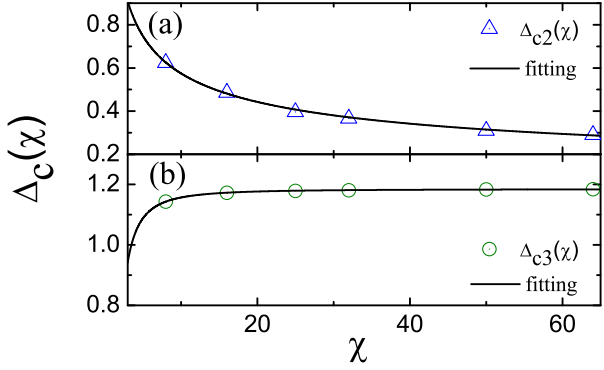}
\caption{(color online)
 Phase transition points $\Delta_{c2}(\chi)$ and $\Delta_{c3}(\chi)$
 as a function of the truncation dimension $\chi$.
 Here, the truncation dimensions are taken as
  $\chi = 8$, $16$, $25$, $32$, $50$, and $64$.
 For the numerical extrapolation,
 the fitting functions are chosen as
 $\Delta_c(\chi)=a+b\chi^{-c}$, where $a$, $b$, and $c$ are a real
 number.
 A best numerical fitting gives
 (a) $a=0.015$, $b=1.366$, and $c=0.387$
  for the XY-Haldane phase transition $\Delta_{c2}(\chi)$, and
 (b) $a=1.185$, $b=-1.748$, and $c=1.800$
   for the Haldane-N\'eel phase transition $\Delta_{c3}(\chi)$.
 The estimated critical points in the thermodynamic limit are given by
  $\Delta_{c2}(\infty)=0.015$ for the XY-Haldane phase transition and
  $\Delta_{c3}(\infty)=1.185$ for the Haldane-N\'eel phase transition.
  The critical points agree well with the results from the previous
  studies \cite{Botet,Sakai,Yajima,Ueda,Nomura}.
 } \label{fig7}
\end{figure}

 %
 Actually, the transition points between the phases depend on
 the truncation dimension $\chi$, i.e.,
 $\Delta_{c2}=\Delta_{c2}(\chi)$ and
 $\Delta_{c3}=\Delta_{c3}(\chi)$.
 As the truncation dimension $\chi$ increases from a lower truncation dimension
 (e.g., $\chi = 8$),
 the phase transitions $\Delta_{c2}(\chi)$ and $\Delta_{c3}(\chi)$ occur
 starting at the lower and the higher values of $\Delta$'s, respectively.
 Then,
 the critical points $\Delta_{c2}(\infty)$ and $\Delta_{c3}(\infty)$
 in the thermodynamic limit can be extrapolated to $\chi \rightarrow \infty$.
 In Fig. \ref{fig7}, we plot the transition points
 (a) $\Delta_{c2}(\chi)$ and (b) $\Delta_{c3}(\chi)$
 as a function of the truncation dimension $\chi$.
 We employ an extrapolation function $\Delta(\chi) = a + b \chi^{-c}$,
 characterized by the fitting constants $a$, $b$, and $c$,
 which guarantees that $\Delta(\infty)$ becomes a
 finite value.
 The numerical fittings give, respectively,
 $a=0.015$,
 $b=1.366$ and $c=0.387$ for the phase transition between the XY
 and Haldane phases and $a=1.185$, $b=-1.748$, and
 $c=1.800$ for the phase transition from the Haldane phase to
 the N\'eel phase.
 In Fig. \ref{fig7}, in the limit of the infinite truncation
 dimension, i.e., $\chi \rightarrow \infty$,
 the fitting function are shown to saturate well to the extrapolated value
 $\Delta(\infty)= a$ which can be regarded as a critical point
 $\Delta_c = \Delta(\infty)$.
 As a result, our extrapolations give the critical points
 $\Delta_{c2}(\infty)= 0.015$ and $\Delta_{c3}(\infty)= 1.185$.
 Our critical points agree well with
 the results $\Delta_{c2}=0.068 \pm 0.003$ from the exact
 diagonalization \cite{Yajima},
 $\Delta_{c2}=-0.01\pm0.03$ from
 the phenomenological renormalization group
 with the finite-size scaling analysis
 \cite{Sakai},
 $\Delta_{c3}=1.17\pm 0.02$ \cite{Botet,Sakai,Nomura}
 and $\Delta_{c3}=1.186$ \cite{Ueda} from the DMRG.

 \section{XY phase}

 The XY phase is known to have a power-law decay of the spin-spin
 correlations with a gapless excitation.
 This implies that there exists no long-range order in the
 XY phase in the thermodynamic limit.
 Actually, in numerical approaches,
 directly characterizing a XY phase from a power-law decay of the spin-spin correlation
 is a non-trivial task.
 This may be the reason why a level spectroscopy of numerical
 approaches has been invented as a useful way to characterize the XY phase.
 However,
 directly detecting a vanishing excitation gap from numerical calculations
 near the transition point is also not a trivial work
 due to a very limited lattice size.
 As discussed in Sec. V, our iMPS approach has verified
 that the transverse string order parameter and the longitudinal
 N\'eel order parameter clearly characterize the Haldane phase and
 the N\'eel phase, respectively, in Figs. \ref{fig6} (a) and (b).
 Obviously, the longitude N\'eel order and all the components of the string order
 become zero in the XY phase even for the finite truncation dimension $\chi$.

 However, such vanishing behaviors of the long-range order parameters in the XY phase
 do not guarantee the existence of a XY phase in the iMPS representation.
 In this sense, it would be worthwhile to discuss
 a practical way to characterize the XY phase in the iMPS
 representation by defining a pseudo order parameter
 for a finite truncation dimension $\chi$.
 Then, some local or non-local properties of our iMPS
 groundstate will be introduced as indicators that
 can be used to distinguish a XY phase from other phases in the iMPS
 representation \cite{Zhou1,Wang1}.

\begin{figure}
\includegraphics[width=3.3in]{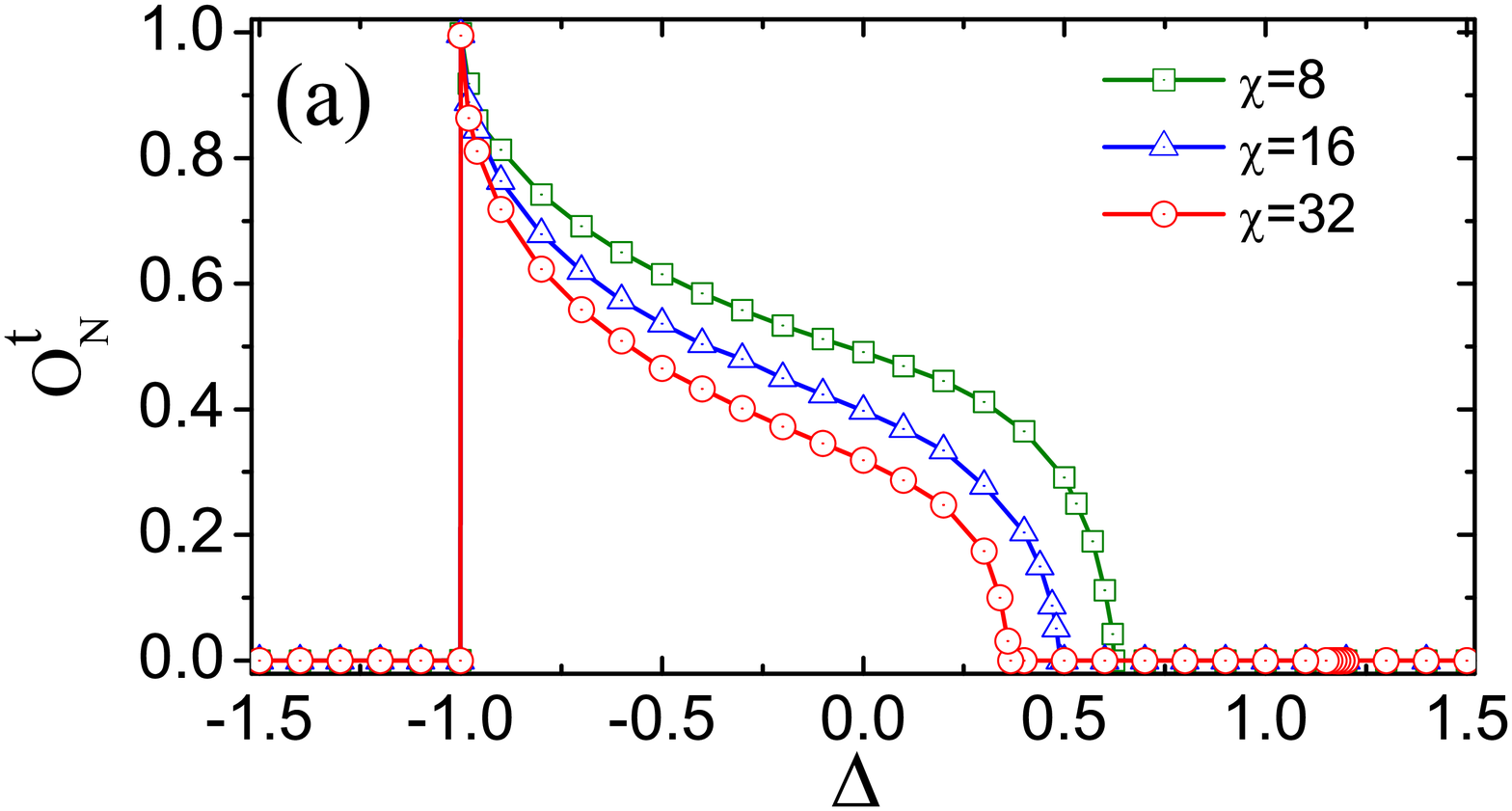}
\includegraphics[width=3.3in]{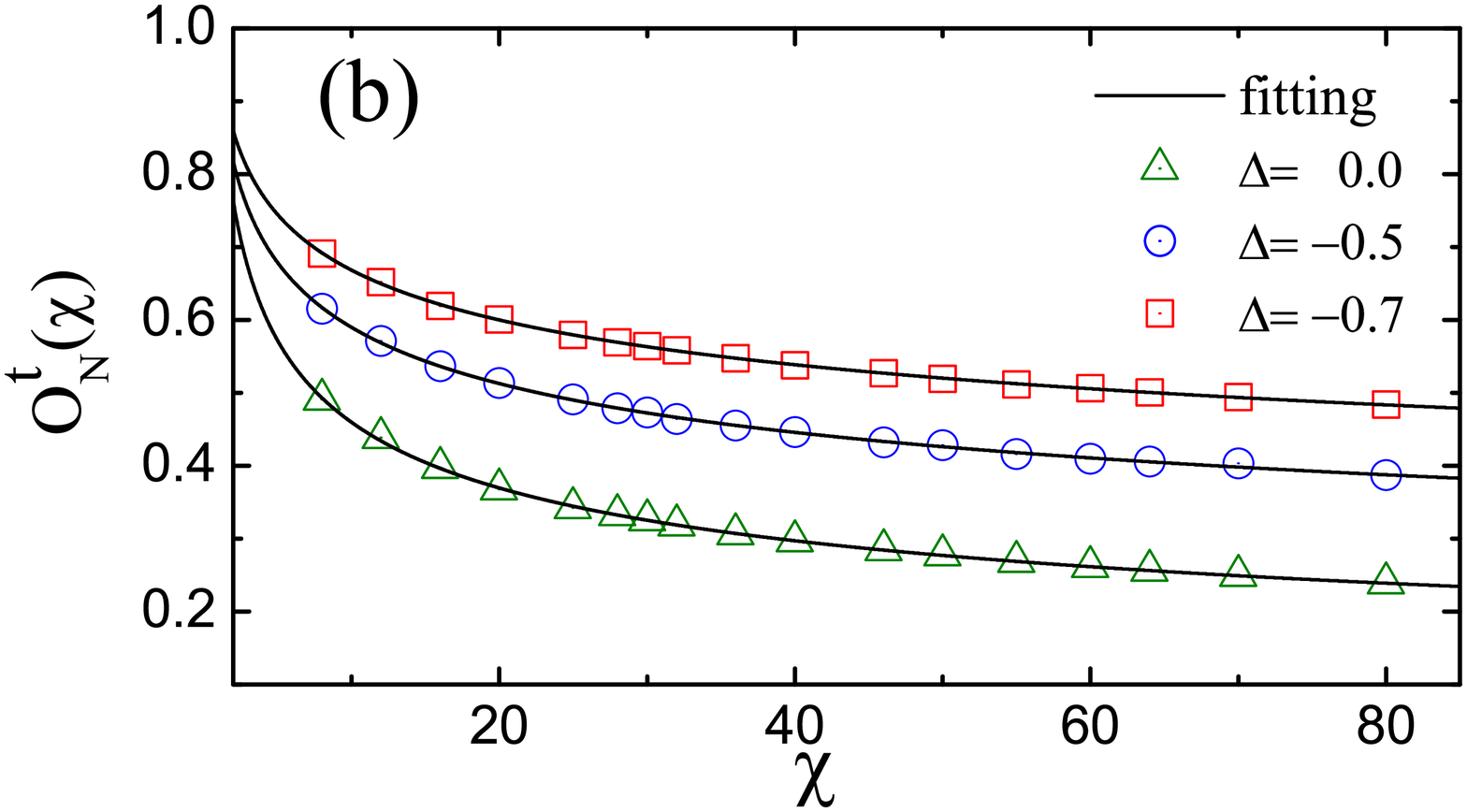}
\caption{(color online)
 (a) Transverse N\'eel order ${O}^{\, t}_{N}={O}^{\, x}_{N} + {O}^{\, y}_{N}$
 as a function of the anisotropic interaction $\Delta$ with
 for various values of the truncation dimension $\chi$.
 Note that, in the XY phase, the transverse N\'eel order is not zero
 for the finite truncation dimensions.
 (b) Transverse N\'eel order $O^{\, t}_{N}(\chi)$
 as a function of the truncation dimension $\chi$
 for  $\Delta = 0.0$, $-0.5$, and $-0.7$.
%
%
 %
 For the numerical extrapolation,
 the fitting functions are chosen as
 ${O}^{\, t}_N(\chi)=a+b\chi^{-c}$, where $a$, $b$, and $c$ are a real
 number.
 A best fitting gives
 the fitting constants as (i) $a=0.446\times 10^{-4}$,
 $b=0.948$, and $c=0.315$ for $\Delta=0.0$,
 (ii) $a=3.643\times 10^{-3}$,
 $b=0.936$, and $c=0.203$ for
 $\Delta=-0.5$, and (iii)
 $a=0.897\times 10^{-4}$, $b=0.956$, and $c=0.156$ for $\Delta=-0.7$.
 These results show that, in the thermodynamic limit, the transverse N\'eel order
 does not exist in the XY phase, i.e., ${O}^{\, t}_N(\chi = \infty) =
 0$.
 } \label{fig8}
\end{figure}

 \subsection{Transverse N\'eel order for finite truncation dimensions}
 Let us consider the transverse N\'eel order in the XY phase.
 In Fig. \ref{fig8} (a), we plot the transverse N\'eel order
 as a function of the anisotropic interaction strength $\Delta$
 for various truncation dimensions.
 Here, the transverse N\'eel order is defined by the sum of
 the $x$- and $y$- components of the N\'eel order, ${O}^{\,
 t}_{N} (\chi)= {O}^{\, x}_{N} (\chi)+{O}^{\, y}_{N}(\chi)$.
 It is shown that the transverse N\'eel order has a finite value
 in the XY phase. Actually, as the truncation
 dimension increases, as discussed in Sec. V,
 the interaction parameter range of
 the XY phase becomes narrower because the transition point $\Delta_{c2}(\chi)$
 between the XY phase and the Haldane phase moves
 to a lower value for a higher truncation dimension.
 Note that, from the transverse N\'eel order,
 the transition points between the XY phase and the Haldane phase
 (non-zero values of the transverse N\'eel order)
 are the same with the values from the string order parameter,
 while the ferromagnetic-XY transition point does not move
 at $\Delta_{c1}=-1.0$.
 Thus, the non-vanishing transverse N\'eel order can be used
 as a pseudo order parameter characterizing the XY phase for
 a finite truncation dimension $\chi$.

 Also, it should be noted that the overall amplitude of the
 transverse N\'eel order becomes smaller as the truncation dimension
 increases in Fig. \ref{fig8} (a).
 In order to understand the transverse N\'eel order in
 the thermodynamic limit, i.e., $\chi \rightarrow \infty$,
 in Fig. \ref{fig8} (b), we plot the transverse N\'eel order
 as a function of the truncation dimension $\chi$
 for, as examples, three anisotropic interaction strengthes $\Delta = 0$,
 $-0.5$, and $-0.7$.
 It is shown clearly that the transverse N\'eel order decreases as
 the truncation dimension increases.
 We perform an extrapolation by especially introducing
 a power-law fitting function ${O}^{\, t}_{N}(\chi)=a+b\chi^{-c}$
 with respect to the truncation dimension $\chi$.
 The numerical fittings give
 (i) $a=0.446\times 10^{-4}$,
 $b=0.948$ and $c=0.315$ for $\Delta=0$,
 (ii)  $a=3.643\times 10^{-3}$,
 $b=0.936$ and $c=0.203$ for $\Delta=-0.5$,
 and (iii)
 $a=0.897\times 10^{-4}$, $b=0.956$ and $c=0.156$ for $\Delta=-0.7$.
 Hence, the extrapolated values of the transverse N\'eel order for
 $\chi \rightarrow \infty$
 are given as ${O}^{\, t}_{N}(\infty) = 0.446\times 10^{-4}$,
 $3.643\times 10^{-3}$, and $0.897\times 10^{-4}$ for $\Delta = 0$,
 $-0.5$, and $-0.7$, respectively.
 This implies that, similar to the power-law decay of spin-spin
 correlations with respect to the lattice distance,
 the transverse N\'eel order follows a power-law decaying to zero with respect
 to the truncation dimension $\chi$.
 These results show that,
 in the XY phase,
 the transverse N\'eel order as well as the longitudinal one
 also becomes zero, ${O}^{\, \alpha}_{N}(\infty) = 0$.
 As a consequence, \textit{both the string and the N\'eel long-range orders
 do not exist in the XY phase.}

 \subsection{ Pseudo local order
             for finite truncation dimensions}
 \begin{figure}
 \includegraphics[width=3.3in]{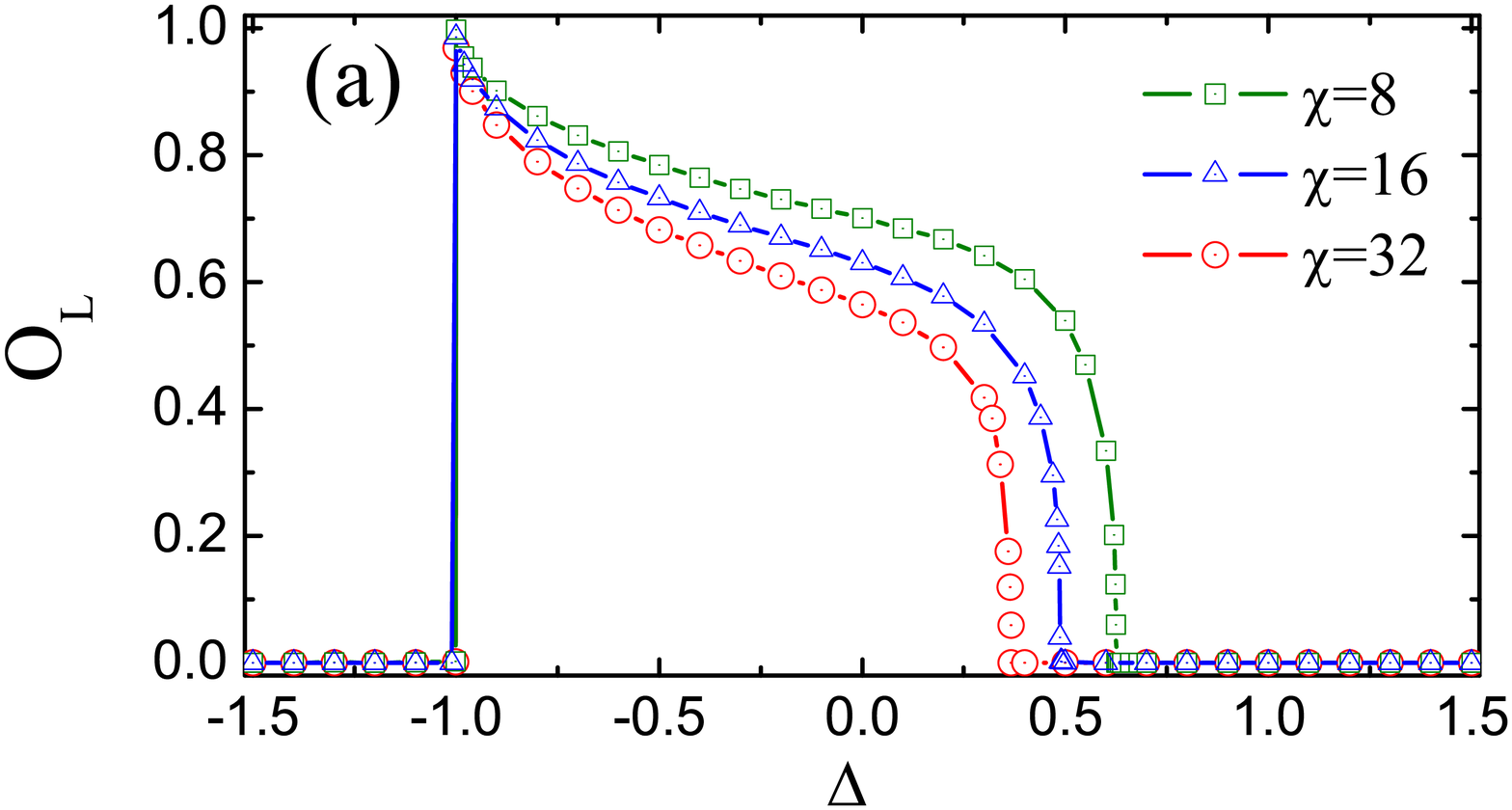}
 \includegraphics[width=3.3in]{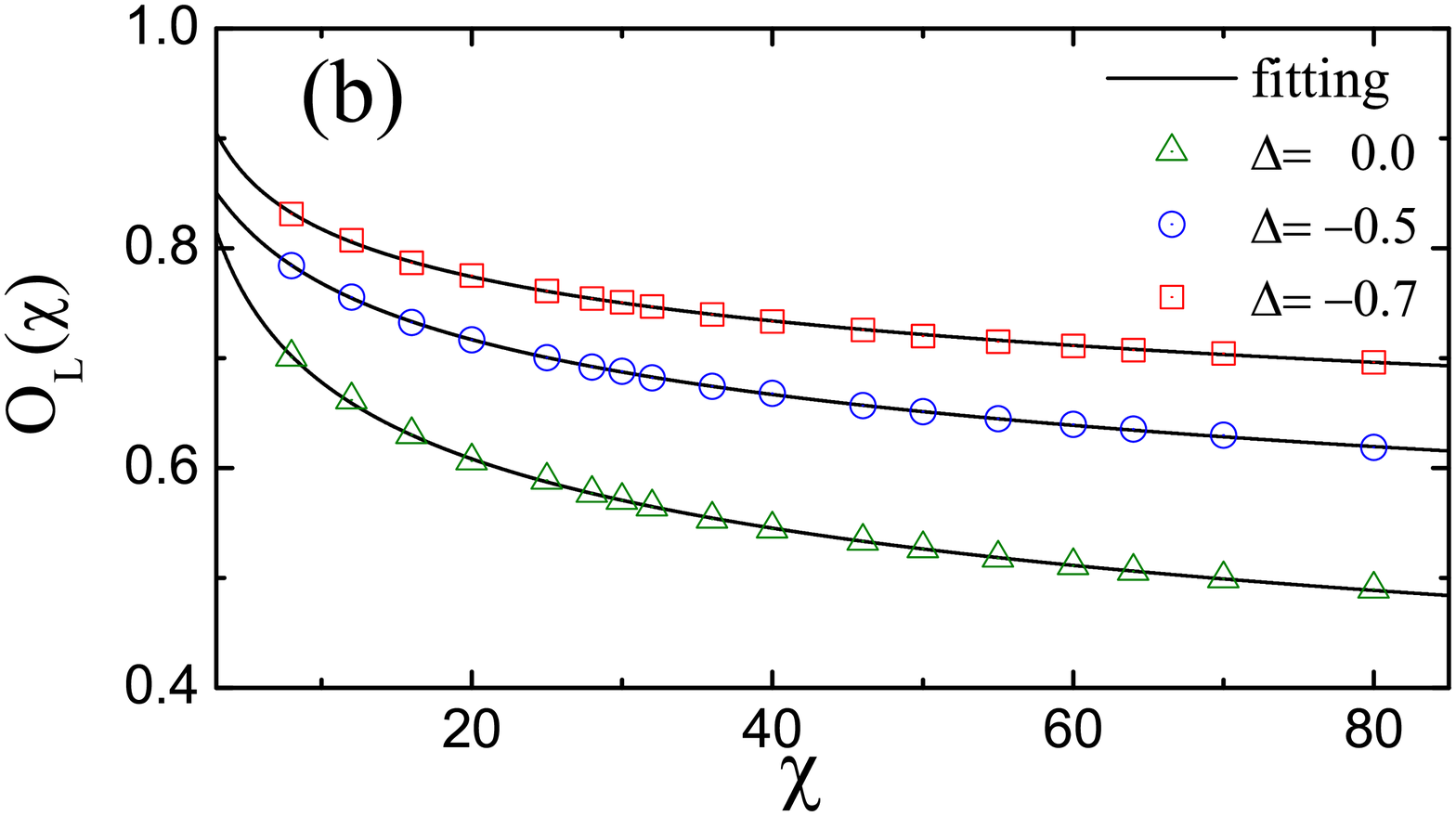}
 \caption{(color online)
 (a) Pseudo local order
 $O_L = \sqrt{\langle S_x \rangle^2 + \langle S_y \rangle^2}$ for
 the XY phase
 as a function of
  the anisotropic interaction $\Delta$ for
 various values of the truncation dimension $\chi$.
  (b) Pseudo local order $O_L(\chi)$
 as a function of the truncation dimension $\chi$
 for  $\Delta = 0.0$, $-0.5$, and $-0.7$.
%
%
 %
 By using the fitting function  ${O}_L(\chi)=a \chi ^{-b} (1+c \chi^{-1})$
 with $a$, $b$, and $c$ being a real
 number,
 the numerical extrapolations give
 the fitting constants as  (i) $a=0.980$,
 $b=0.159$, and $c=-0.028$ for $\Delta=0.0$,
 (ii) $a=0.998$, $b=0.108$, and $c=-0.120$ for
 $\Delta=-0.5$, and (iii)
 $a=0.967$, $b=0.075$, and $c=0.048$ for $\Delta=-0.7$.
 } \label{fig9}
 \end{figure}
 Recently,
 a pseudo local order have been suggested for the
 XY phase in the iMPS representation in Ref. \onlinecite{Wang1}.
 The local order can be defined as
 ${\cal O}_L = \sqrt{\langle S_x \rangle^2 + \langle S_y \rangle^2}$.
 In Fig. \ref{fig9} (a), we plot the pseudo local order
 ${\cal O}_L$ as a function of the anisotropic interaction strength $\Delta$
 for various truncation dimensions.
 It is shown that the defined local order has a finite value
 in only the XY phase.
 Similar to the transverse N\'eel order,
 from the defined local order,
 the transition points between the XY phase and the Haldane phase
 (non-zero values of the defined local order)
 are detected at the same values from the string order parameter
 for the same truncation dimension $\chi$.
 Also,
 the ferromagnetic-XY transition points from the defined local order
 does not move
 at $\Delta_{c1}=-1.0$.
 Compared with the transverse N\'eel order,
 Fig. \ref{fig9} (a)
 shows that the overall amplitude of the defined local order becomes smaller
 as the truncation dimension $\chi$ increases.
 In Fig. \ref{fig9} (b), we plot the transverse N\'eel order
 as a function of the truncation dimension $\chi$
 for, as examples, three anisotropic interaction strengthes $\Delta = 0$,
 $-0.5$, and $-0.7$.
 It is shown clearly that the defined local order decreases as
 the truncation dimension increases.
 We perform an extrapolation  with respect to the truncation dimension $\chi$
 by using the same fitting function
 ${O}_L(\chi)=a \chi ^{-b} (1+c \chi^{-1})$,
 with $a$, $b$, and $c$ being a real
 number, given in Ref.
 \onlinecite{Wang1}.
 Figure \ref{fig9} (b) shows
 the behaviors of the defined local order for the XY phase
 agree well with the results of spin-1/2 XXZ model in Ref.
 \onlinecite{Wang1}.
 Hence, similar to the transverse N\'eel order,
 the defined local order can be used
 as a pseudo order parameter characterizing the XY phase for
 a finite truncation dimension $\chi$.

\section{Entanglement entropy,
         Central charge, and universality class for phase transitions}

 Instead of using order parameters, recently,
 various types of quantum entanglement measures have been proposed as
 an indicator characterizing quantum phase transitions \cite{QPT,Skr}.
 One of successful measures is the von Neumann entropy for a
 bipartite system \cite{Osborne,Chung,Cho,Vidal3}.
 Singular behaviors of bipartite entanglements for a pure state
 reveal quantum critical behaviors,
 which has been verified as being universal
 by extensive studies in many one-dimensional systems \cite{Osterloh,entropy}.
%
%
\begin{figure}
\includegraphics[width=3.3in]{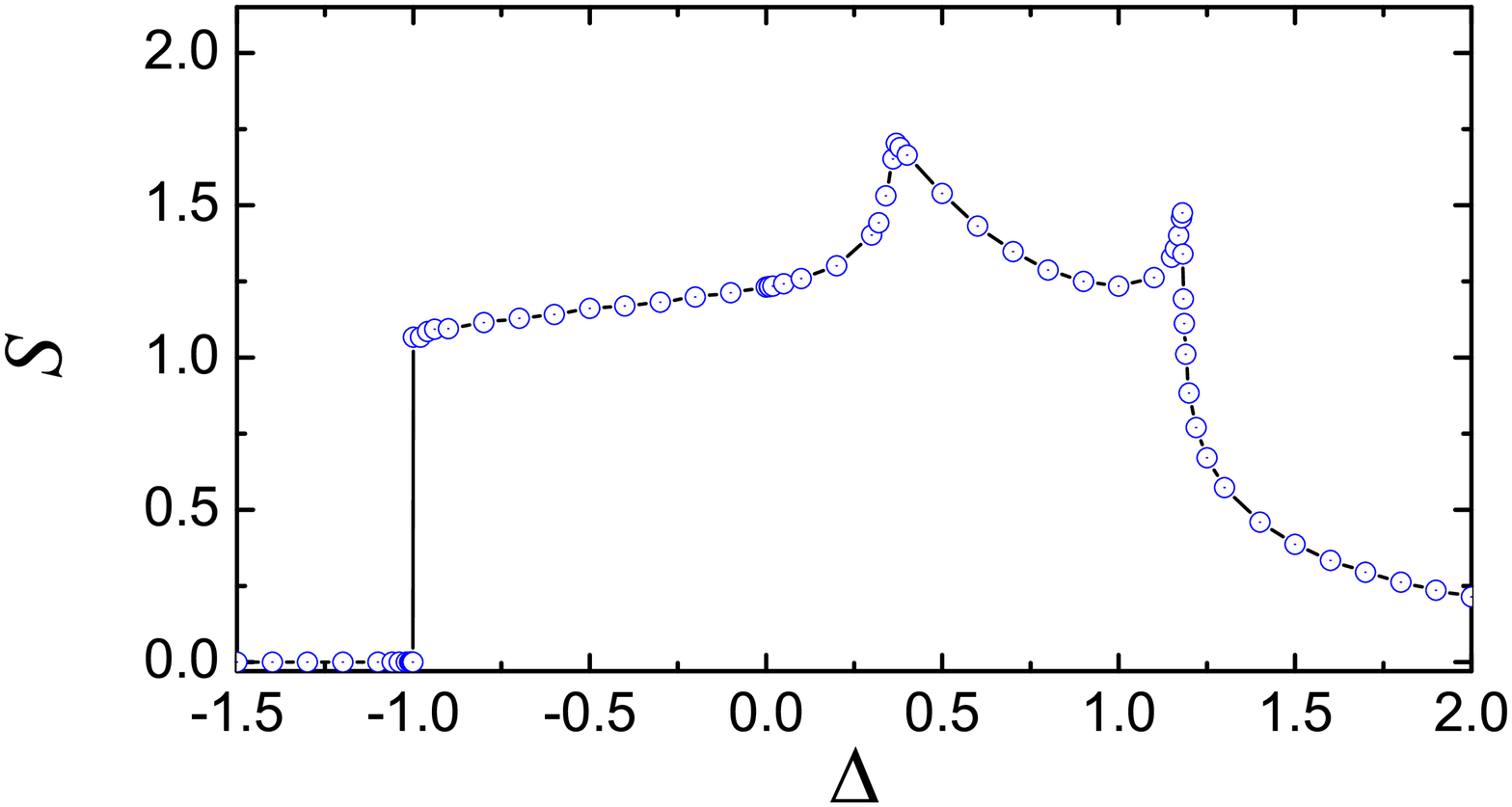}
\caption{(color online)
 Von Neumann entropy as a function of the anisotropic interaction $\Delta$ for
 the truncation dimension $\chi=32$.
 The transition points are seen at $\Delta_{c1}=-1.0$, $\Delta_{c2} = 0.366$,
 and $\Delta_{c3}=1.180$.
 This result is consistent with the phase transition points from the order
 parameters in Fig. \ref{fig6}.
 }
 \label{fig10}
\end{figure}

 {\it Von Neumann entropy singularities.}$-$
 In the iMPS approach, the von Neuman entropy can be explored.
 Let us recall the diagonal matrix $\lambda$.
 As discussed in Sec. II,
 the elements of the diagonal matrix $\lambda^{[i]}_{\alpha_i}$
 are the Schmidt decomposition coefficients of the bipartition
 between the semi-infinite chains $L(-\infty,...,i)$ and
 $R(i+1,...,\infty)$.
 This implies that Eq. (\ref{wave}) can be rewritten by
 $|\Psi\rangle = \sum^\chi_{\alpha=1} \lambda_{\alpha}
 |\psi^L_\alpha\rangle |\psi^R_\alpha \rangle$,
 where $|\psi^L_\alpha \rangle$ and $|\psi^R_\alpha \rangle$ are the Schmidt bases
 for the semi-infinite chains $L(-\infty,...,i)$ and
 $R(i+1,...,\infty)$, respectively.
 For the bipartition, then,
 the von Neumann entropy $S$ can be defined as \cite{Bennett}
 $S= - {\rm Tr} [\varrho_L \log
 \varrho_L] = - {\rm Tr} [\varrho_R \log \varrho_R]$, where $\varrho_L =
 {\rm Tr}_R\,  \varrho$ and $\varrho_R={\rm Tr}_L\,  \varrho$ are the reduced density
 matrices of the subsystems $L$ and $R$, respectively, with the density
 matrix $\varrho=|\Psi \rangle \langle \Psi|$.
 For the semi-infinite chains $L$ and $R$ in the iMPS representation,
 the von Neumann entropy $S$ is given by
 \begin{equation}
  S = -\sum_{\alpha=1}^{\chi} \lambda^2_\alpha \log
  \lambda^2_\alpha.
  \label{entropy}
 \end{equation}
 In Fig. \ref{fig10}, we plot the von Neumann entropy
 as a function of $\Delta$ for $\chi =32 $.
 In the entropy, there are three singular points that consist
 of two local peaks ( $\Delta =0.366$ and $\Delta =1.180$, respectively )
 and one discontinuous point ($\Delta=-1.0$).
 In fact, the singular points correspond to
 the transition points from the string order parameters
 and the N\'eel order parameter.
 It is shown that the von Neumann entropy
 captures the phase transitions.
 The discontinuity of the von Neumann entropy indicates
 that a discontinuous phase transition occurs between the
 Ferromagnetic phase and the XY phase.
 The two singular peaks show that the XY-Haldane phase transition
 and the Haldane-N\'eel phase transition belong to
 a continues phase transition.
 Especially, it should be noted that,
 in our iMPS representation, the von Neumann entropy can
 detect the BKT phase transition between the XY phase and the Haldane phase.

\begin{figure}
\includegraphics[width=3.3in]{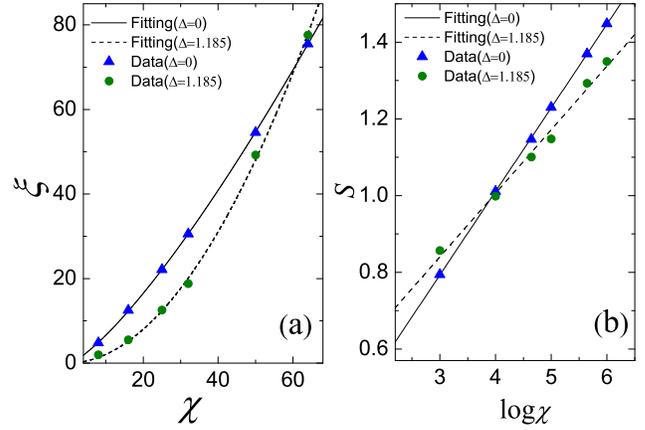}
\caption{(color online)
 (a) Correlation length $\xi$ as a function of the truncation
 dimension $\chi$ at the critical points $\Delta_{c2}=0$
 and $\Delta_{c3}=1.185$, respectively.
 The power curve fittings $\xi = \eta \chi^{\, \kappa}$ yields
 $\eta=0.330$ and
 $\kappa=1.306$ at $\Delta_{c2}=0$, and
 $\eta=0.023$ and $\kappa=1.955$ at $\Delta_{c3}=1.185$, respectively.
  (b) Scaling of the von Neumann entropy $S$
  with respect to the truncation dimension $\chi$
  at the critical points.
  From the fitting function $S(\chi) =a + b \log \chi$,
  the fitting constants are given as $a= 0.139$ and $b=0.218$
  at $\Delta_{c2}=0$, and  $a=0.343$ and $b=0.166$ at $\Delta_{c3}=1.185$.
 From (i) $\kappa=1.306$ and $b=0.218$ at $\Delta_{c2}=0.0$,
  and (ii) $\kappa=1.955$ and $b=0.166$ at $\Delta_{c3}=1.185$ in (a) and (b),
 the central charges are determined as (i) $c=1.001$ and (ii) $c=0.509$, respectively.
 } \label{fig11}
\end{figure}

 {\it Central charge and universality class.}$-$
 For one-dimensional quantum spin models, in
 general, the logarithmic scaling of von Neumann entropy
 was conformed to exhibit conformal invariance \cite{Korepin}
 and the scaling is governed by a
 universal factor, i.e., a central charge of the
 associated conformal field theory.
 In fact, in the iMPS representation, a diverging entanglement at quantum criticality
 gives simple scaling relations for (i) the von Neumann entropy
  $S$ and (ii) a correlation length $\xi$ with respect to
 the truncation dimension $\chi$ as \cite{Korepin,Tagliacozzo,Pollmann}
\begin{subequations}
 \begin{eqnarray}
  S &\sim& \frac {c \kappa} {6} \log \chi,
  \label{S}
 \\
  \xi &\sim & \eta\chi^{\,\kappa},
 \label{xi}
\end{eqnarray}
\end{subequations} where $c$ is a central charge and $\kappa$ is
 a so-called finite-entanglement scaling exponent.
 Here, $\eta$ is a constant.
 By using Eqs. (\ref{S}) and (\ref{xi}), then, a central charge can be
 obtained numerically at a critical point.

 In the iMPS approach,
 the correlation length $\xi$ can be obtained from the transfer matrix $T$
 defined in Fig. \ref{fig1} (c).
 Actually, for a given $\chi$, the finite correlation length in the iMPS representation
 can be defined as
 $\xi(\chi) = 1/ \log {|\mu_0(\chi)/\mu_1(\chi)|}$,
 where the $\mu_0$ and $\mu_1$ are the largest
 and the second largest eigenvalues
 of the transfer matrix $T$, respectively.
 In Fig. \ref{fig11}, we plot
 (a) the correlation length $\xi$ and (b) the von Neumann entropy
 as a function of the truncation
 dimension $\chi$ at the critical points $\Delta_{c2} =0.0 $ and $\Delta_{c3}=1.185$.
 Here, the truncation dimensions are taken as $\chi=8$, $16$, $25$,
 $32$, $50$, and $64$.
 It is shown that both the correlation length $\xi$ and the von
 Neumann entropy $S$ diverge as the truncation dimension $\chi$ increases.
 In order to obtain the central charges,
 we use the numerical fitting functions, i.e.,
 $S(\chi) = a + b \log \chi$ and $\xi(\chi) = \eta \chi^{\,\kappa}$.
 From the numerical fittings of the von Neumann entropies $S$,
 the fitting constants are given as
 $a=0.139$ and $b=0.218$ for $\Delta_{c2}=0$, and
 $a=0.343$ and $b=0.166$ for $\Delta_{c3}=1.185$.
 Also, the power-law fittings on the correlation lengthes $\xi$
 give the numerical fitting constants as
 $\kappa=1.306$ and $\eta=0.330$ for $\Delta_{c2}=0$, and
 $\kappa=1.955$ and $\eta=0.023$ for $\Delta_{c3}=1.185$.
 As a result,
 the central charges are given by
 $c=1.001$  for $\Delta_{c2}=0$ and $c=0.509$ for
 $\Delta_{c3}=1.185$.
 Our central charges are very close to
 the exact values $c=1$ and $c=0.5$, respectively.
 Therefore,
 the XY-Haldane  phase transition at $\Delta_{c2}=0$
 belongs to the Heisenberg universality class,
 while the Haldane-N\'eel phase transition at $\Delta_{c2}=1.185$
 belongs to the two-dimensional classical Ising universality class.

\begin{figure}
\includegraphics[width=3.3in]{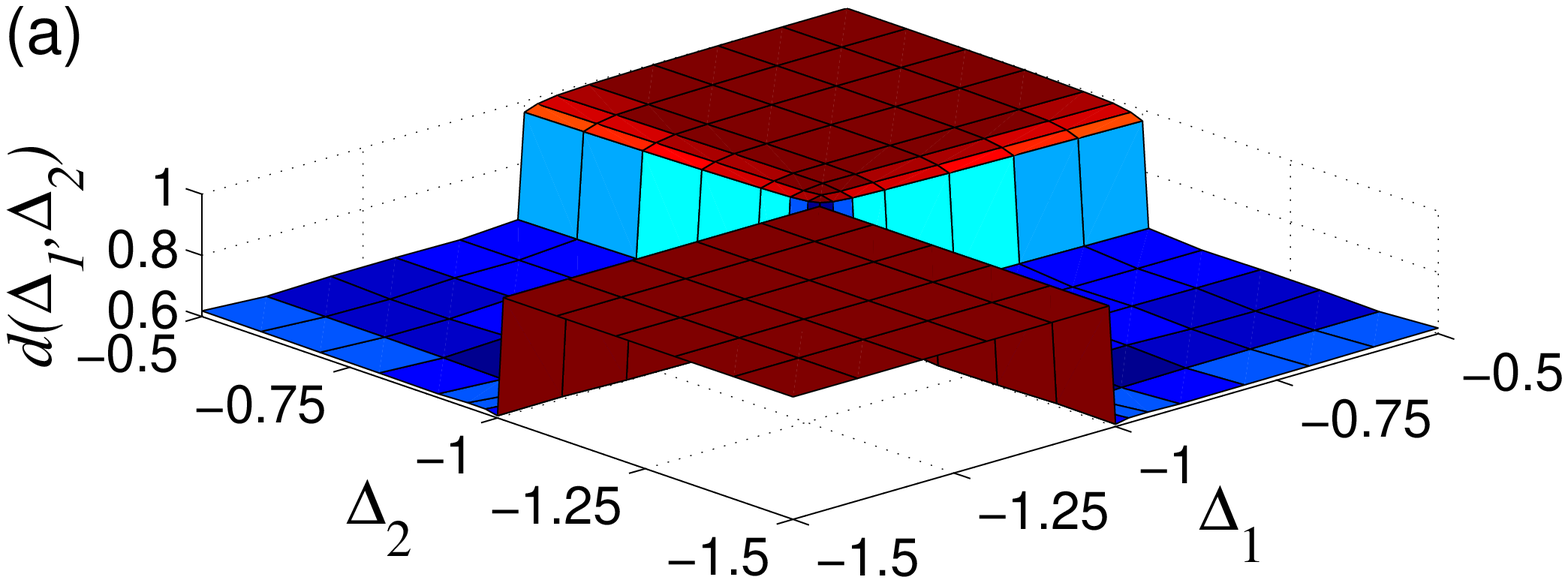}
\includegraphics[width=3.3in]{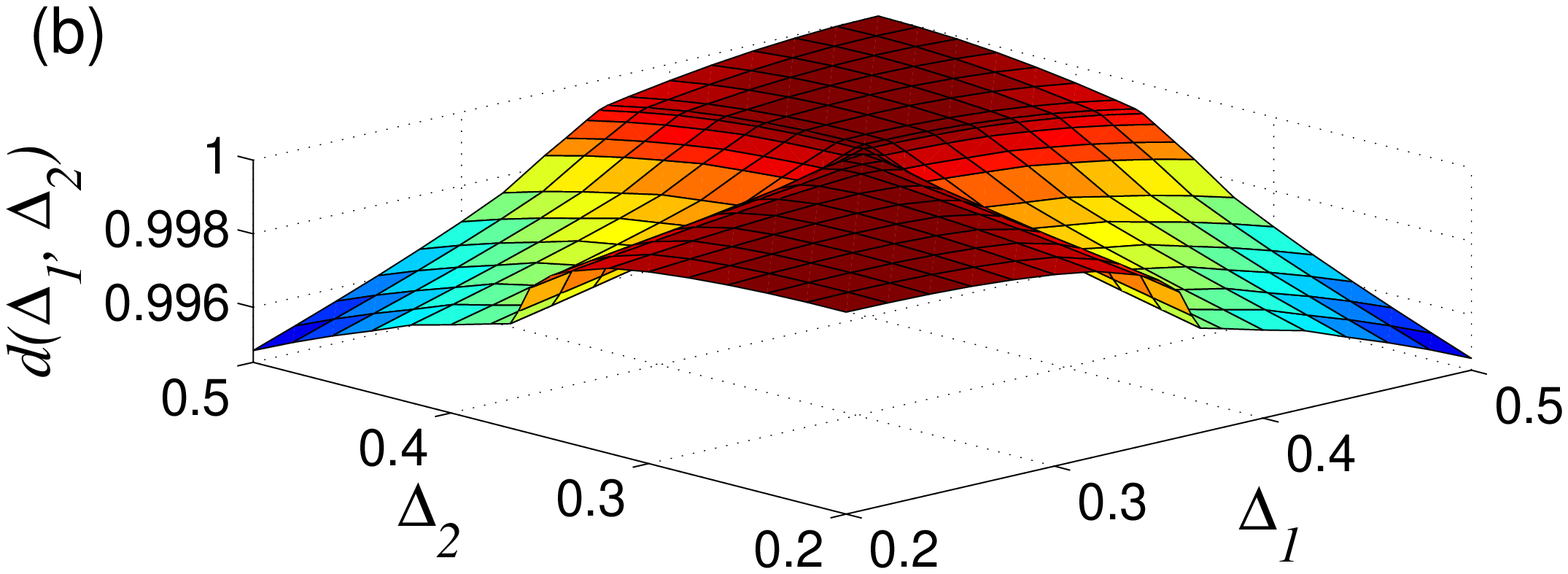}
\includegraphics[width=3.3in]{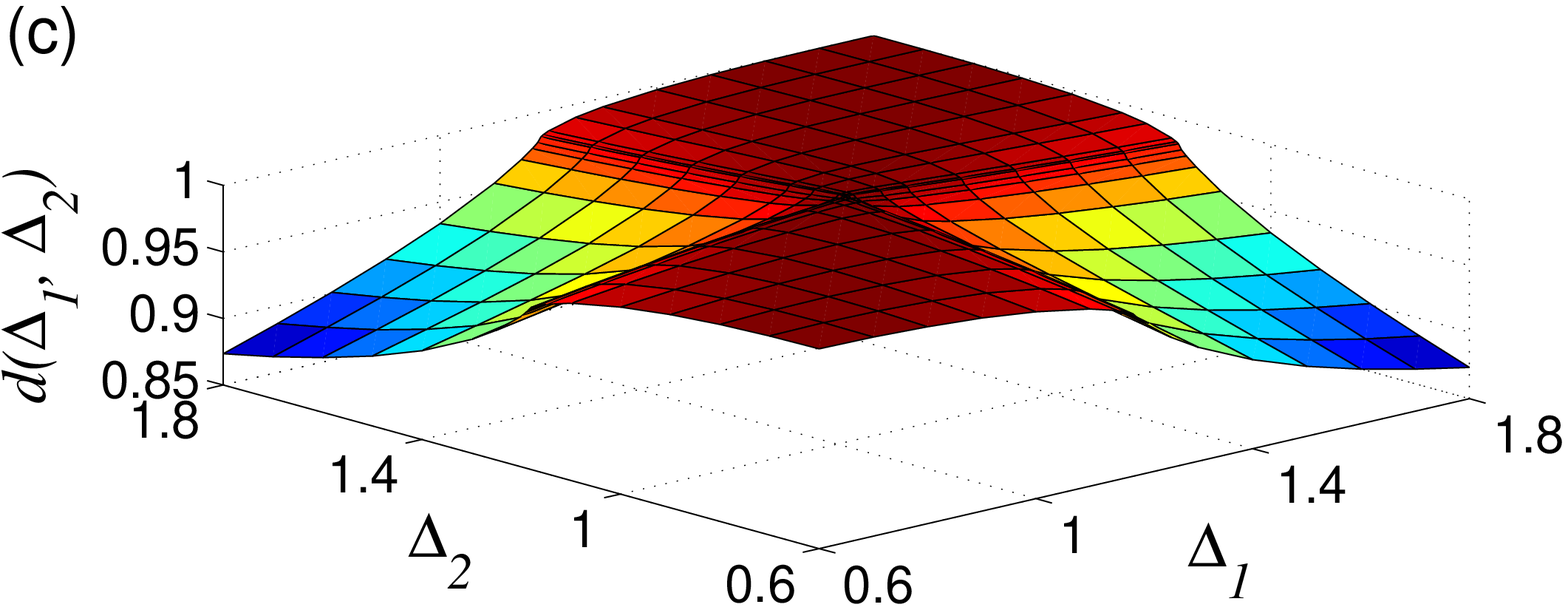}
\caption{(color online)
 Fidelity per lattice site (FLS)
 $d(\Delta_1, \Delta_2)$ as a function of the anisotropic
 interactions
 $\Delta_1$ and $\Delta_2$ for the truncation dimension $\chi = 32$.
 The pinch points are seen at (a) $\Delta = -1.0$,
 (b) $\Delta = 0.366$, and (c) $\Delta = 1.180$.
 These pinch points in the FLS
 are consistent with the phase transition points from the order
 parameters
 in Fig. \ref{fig6}.
 In (a) for the ferromagnetic-XY phase transition,
 the FLS discontinuity indicates that
 a first-order phase transition occurs at the transition point $\Delta=-1.0$.
 In (b) for the XY-Haldane phase transition
 and (c) for the Haldane-N\'eel phase transition,
 the continuous behaviors of the FLS across
 the pinch points implies that a continuous phase transition occurs
 at the transition points.
 It should be noted that the FLS captures a BKT phase
 transition between the XY and Haldane phases in (b).}
\label{fig12}
\end{figure}

\section{Fidelity per lattice site for phase transitions}

 For a quantum phase transition,
 the groundstate of a system undergoes a drastic change
 in its structure at a critical point \cite{Zanardi,Zhou2,Rams}.
 In fact,
 the groundstates in different phases
 should be orthogonal because the states are distinguishable
 in the thermodynamic limit \cite{Zhao1,Zhou2}.
 It implies that a comparison between quantum many-body states in different phases
 can signal quantum phase transitions
 regardless of what type of internal order exists in
 the states.
 Thus, as an alternative way to explore quantum phase transitions,
 the groundstate fidelity has been used in the last few years
 \cite{Zanardi,Zhou2,Gu3,Rams,Gong,Zhao2,Mukherjee,Liu,Gu2,Xiao,Chen}.
 In contrast to quantum entanglement,
 the fidelity is a measure of similarity between two states.
 An abrupt change of the fidelity can then be expected across a critical
 point (in the thermodynamic limit).
 Based on understanding such a property of the groundstate fidelity
 near critical points,
 several measures have been suggested such as FLS \cite{Zhou2},
 reduced fidelity \cite{Liu},
 fidelity susceptibility \cite{Gu3}, density-functional fidelity \cite{Gu2},
 and operator fidelity \cite{Xiao}.
 However, it is known that the fidelity susceptibility cannot detect a
 BKT type phase transition \cite{Ren,Chen}.
 Thus, the fidelity approaches have been thought to be a model-dependent
 indicator for quantum phase transitions.
 In order to show that a BKT type phase transition can be captured
 by the FLS approach, therefore,
 we discuss the FLS in the iMPS representation in this section.

 Once one obtains the groundstate as a function of the anisotropic
 interaction strength $\Delta$,
 the groundstate fidelity is defined  as
 $F(\Delta_1,\Delta_2)=|\langle \psi(\Delta_2)|\psi(\Delta_1)\rangle|$.
 Following Ref. \onlinecite{Zhou3}, we define the groundstate FLS as
\begin{equation}
 \ln d(\Delta_1,\Delta_2) \equiv \lim_{L \rightarrow \infty}
  \frac{\ln F(\Delta_1,\Delta_2)}{L},
\end{equation}
 where $L$ is the system size.
 The FLS is well defined in the thermodynamic limit even if
 $F(\Delta_1,\Delta_2)$ becomes trivially zero.
 From the fidelity $F(\Delta_1,\Delta_2)$,
 the FLS has several properties as
 (i) normalization $d(\Delta,\Delta)=1$, (ii) symmetry
 $d(\Delta_1,\Delta_2)=d(\Delta_2,\Delta_1)$, and (iii) range $0 \le
 d(\Delta_1,\Delta_2)\le 1$.
 Within the iMPS approach,
 the FLS $d(\Delta_1,\Delta_2)$ \cite{Zhou3} is given by the largest eigenvalue $\mu_0$ of
 the transfer matrix $T$ up to the corrections that decay exponentially
 in the linear system size $L$. Then, for the infinite-size system,
 $d(\Delta_1,\Delta_2)= \mu_0$.

 In Fig. \ref{fig12}, the groundstate FLSs $d$
 are displayed as a function of the anisotropic interaction
 parameters ($\Delta_1$,$\Delta_2$) with the truncation dimension
 $\chi = 32$.
 In the FLS surfaces, it is shown that there are three pinch points
 in the spin-1 XXZ model.
 Each pinch points correspond to the transition points identified from the order
 parameters.
 In Fig. \ref{fig12} (a),
 the fidelity undergoes an abrupt change, which means that the first-order
 phase transition between the Ferromagnetic phase and the XY phase
 occurs at the pinch point.
 It is consistent with the discontinuous entropy in Fig. \ref{fig10}.
 In Fig. \ref{fig12} (b), the pinch point corresponds to
 the XY-Haldane phase transition point.
 Contrasted to the fidelity susceptibility,
 the FLS is able to detect a BKT transition successfully,
 which is consistent with the result of our von Neumann entropy.
 Figure \ref{fig12} (c) shows another continuous phase transition, i.e.,
 the Haldane-N\'eel phase transition.
 Hence, it is shown that {\it the FLS approach can
 be applied to characterize quantum phase transitions as a universal
 indicator \cite{Zhou2,Zhou3}.}

\section{Conclusions and remarks}

 We have investigated the string correlations in an infinite-size lattice of
 a spin-1 XXZ chain. In order to obtain a LRSO directly
 rather than an extrapolated string
 order in a finite-size system,
 the iMPS presentation has been employed
 and the groundstate wavefunction of the infinite lattice system
 has been numerically generated
 by the iTEBD algorithm.
 It was shown that the $x$- and
 $y$-components of the string
 correlations decay exponentially in the N\'eel phase, while they show
 a unique behavior of two-step decaying to zero within a relatively very large lattice distance
 in the XY phase. That is, there is no long-range transverse string order
 in the XY phase and the N\'eel phase.
 However, in the Haldane phase,
 the string correlations are saturated
 to finite values for a relatively smaller lattice distance,
 which shows clearly the existence of a LRSO.
 Consistently, the N\'eel order does not exit in both the XY phase
 and the Haldane phase.
 This result verifies that both the $x$- and
 $y$-components of the LRSO
 are the order parameters characterizing the Haldane phase.
%
%
 The estimated critical points agree well with
 the previous results as
 $\Delta_{c2} = 0$ for the XY-Haldane phase transition
  and $\Delta_{c3} = 1.185$ for the Haldane-N\'eel phase transition.

 Further, the behaviors of the von Neumann entropy and the FLS
 have been discussed at the phase transition points.
 Both the von Neumann entropy and the FLS
 capture the corresponding phase transition points including the BKT point,
 which is consistent with the results
 from the string order parameter.
 Consequently, the von Neumann entropy as well as
 the fidelity approach based on the FLS can
 be applied to characterize quantum phase transitions as a universal
 phase transition indicator.
 Moreover, from a finite-entanglement scaling of the von Neumann entropy
 with respect to  the truncation dimension,
 the central charges are obtained as
 $c \simeq 1$ at  $\Delta_{c2} = 0$ and $c \simeq 0.5$ at $\Delta_{c3} = 1.185$,
 respectively, which shows
 the XY-Haldane phase transition at $\Delta_{c2}=0$
 belongs to the Heisenberg universality class
 while the Haldane-N\'eel phase transition at $\Delta_{c2}=1.185$
 belongs to the two-dimensional classical Ising universality class.

 Contrary to other approaches,
 a feature of the iMPS approach is that, just from the iMPS
 groundstate,
 its critical behavior can be captured
 irrespective of whether a system has a finite excitation energy gap or not
 because, in principle, local and non-local order parameters can be calculated directly.
 Furthermore,
 von Neumann entropy and FLS can be used
 as a universal phase transition indicator for quantum phase transition in the iMPS representation.
 Hence,
 this iMPS approach would be widely applicable for capturing quantum critical
 phenomena in one-dimensional lattice many-body systems.

\acknowledgements
 This work was supported by the Fundamental Research Funds for
 Central Universities (Project Nos. CDJZR10100027, CDJXS11102213, and CDJXS11102214)
 and the National Natural Science Foundation of
 China (Grant Nos. 10874252 and 11174375).

\end{document}